\documentclass[floatfix,twocolumn,prl,amsmath]{revtex4-2}
\usepackage{graphicx}
\usepackage{bm}
\usepackage{amssymb}
\usepackage{dcolumn}
\usepackage{subfigure}
\usepackage{multirow}
\usepackage{mathrsfs}
\DeclareMathAlphabet{\mathscrbf}{OMS}{mdugm}{b}{n}
\usepackage{color}

\usepackage{threeparttable}
\begin{document}
\newcommand{\intqspa}{\int\!\!\frac{\rmd^d q}{(2\pi)^d}}
\newcommand{\intqspathr}{\int\!\!\frac{\rmd^3 q}{(2\pi)^3}}
\newcommand{\intqspatwo}{\int\!\!\frac{\rmd^2 q}{(2\pi)^2}}
\newcommand{\intkspatwo}{\int\!\!\frac{\rmd^2 k}{(2\pi)^2}}
\newcommand{\intkspa}{\int\!\!\frac{\rmd^d k}{(2\pi)^d}}
\newcommand{\intkspapri}{\int\!\!\frac{\rmd^d k'}{(2\pi)^d}}
\newcommand{\vn}[1]{{\boldsymbol{#1}}}
\newcommand{\vht}[1]{{\boldsymbol{#1}}}
\newcommand{\matn}[1]{{\bf{#1}}}
\newcommand{\matnht}[1]{{\boldsymbol{#1}}}
\newcommand{\bege}{\begin{equation}}
\newcommand{\gretke}{G_{\vn{k} }^{\rm R}(\mathcal{E})}
\newcommand{\gret}{G^{\rm R}}
\newcommand{\gadv}{G^{\rm A}}
\newcommand{\gmat}{G^{\rm M}}
\newcommand{\gles}{G^{<}}
\newcommand{\ghat}{\hat{G}}
\newcommand{\sigmahat}{\hat{\Sigma}}
\newcommand{\glesone}{G^{<,{\rm I}}}
\newcommand{\glestwo}{G^{<,{\rm II}}}
\newcommand{\gspec}{G^{\rm S}}
\newcommand{\glesthree}{G^{<,{\rm III}}}
\newcommand{\magdir}{\hat{\vn{n}}}
\newcommand{\sigmaret}{\Sigma^{\rm R}}
\newcommand{\sigmales}{\Sigma^{<}}
\newcommand{\sigmalesone}{\Sigma^{<,{\rm I}}}
\newcommand{\sigmalestwo}{\Sigma^{<,{\rm II}}}
\newcommand{\sigmalesthree}{\Sigma^{<,{\rm III}}}
\newcommand{\sigmaadv}{\Sigma^{A}}
\newcommand{\ee}{\end{equation}}
\newcommand{\bal}{\begin{aligned}}
\newcommand{\defbar}{\overline}
\newcommand{\SM}{\scriptstyle}
\newcommand{\rmd}{{\rm d}}
\newcommand{\rme}{{\rm e}}
\newcommand{\eal}{\end{aligned}}
\newcommand{\crea}[1]{{c_{#1}^{\dagger}}}
\newcommand{\annihi}[1]{{c_{#1}^{\phantom{\dagger}}}}
\newcommand{\udot}{\overset{.}{u}}
\newcommand{\exponential}[1]{{\exp(#1)}}
\newcommand{\phandot}[1]{\overset{\phantom{.}}{#1}}
\newcommand{\phandag}{\phantom{\dagger}}
\newcommand{\Trace}{\text{Tr}}
\setcounter{secnumdepth}{2}
\title{Bandgaps of insulators from moment-functional based spectral density-functional theory}
\author{Frank Freimuth$^{1,2}$}
\email[Corresp.~author:~]{f.freimuth@fz-juelich.de}
\author{Stefan Bl\"ugel$^{1}$}
\author{Yuriy Mokrousov$^{1,2}$}
\affiliation{$^1$Peter Gr\"unberg Institut and Institute for Advanced Simulation,
Forschungszentrum J\"ulich and JARA, 52425 J\"ulich, Germany}
\affiliation{$^2$ Institute of Physics, Johannes Gutenberg University Mainz, 55099 Mainz, Germany
}
\begin{abstract}
  Within the method of spectral moments it is possible to construct the spectral function
  of a many-electron system
  from the first $2P$ spectral moments ($P=1,2,3,\dots$).
  The case $P=1$ corresponds to standard Kohn-Sham density functional theory (KS-DFT).
  Taking $P>1$ allows us to consider additional important properties of the uniform electron
  gas (UEG) in the
  construction of suitable moment potentials
  for moment-functional based spectral density-functional theory (MFbSDFT).
  For example, the
  quasiparticle renormalization factor $Z$, which is not explicitly considered in KS-DFT,
  can be included easily. In the 4-pole approximation of the spectral function of the
  UEG (corresponding to $P=4$) we can reproduce the momentum distribution,
  the second spectral moment, and the charge response acceptably well, while a treatment of
  the UEG by KS-DFT reproduces from these properties only the charge response.
  For weakly and moderately correlated systems we may reproduce the most important aspects
  of the 4-pole approximation by an optimized two-pole model, which leaves out
  the low-energy
  satellite band. From the optimized two-pole model we
  extract \textit{parameter-free universal} moment potentials for MFbSDFT, which
  improve the description of the bandgaps in Si, SiC, BN, MgO, CaO, and ZnO significantly.
\end{abstract}
\maketitle
\section{Introduction}
Density functionals for KS-DFT are often constructed from the
exchange-correlation energy of the UEG~\cite{vwn,PhysRevB.45.13244}.
Several other important properties of the UEG, such as the
quasiparticle renormalization factor $Z$~\cite{Haule2022_diagrammatic},
the effective mass $m^{*}$, the Landau liquid parameters,
the momentum distribution function $n_k$~\cite{PhysRevB.66.235116},
and the spectral moments~\cite{PhysRevB.69.045113}
are not built-in explicitly 
into KS-DFT. In particular, KS-DFT uses $Z=1$ by construction.
In order to obtain realistic $Z$ factors 
KS-DFT may be combined with many-body techniques
such as DMFT~\cite{rmp_dmft,RevModPhys.78.865}.
Within DMFT the effective mass enhancement is correlated with the
inverse quasiparticle renormalization
factor, i.e., $m^{*}/m=Z^{-1}$~\cite{PhysRevB.99.245128}, and
consequently DMFT predicts
many-body corrections of this quantity as well.

There are several reasons why standard KS-DFT uses only the exchange-correlation
energy of the UEG explicitly, while not employing additionally any other of its many 
well-studied properties directly. The most important reason is that the Hohenberg-Kohn theorem
establishes a direct relation between the exchange-correlation energy of the UEG
and the one of the real solid studied by
KS-DFT~\cite{PhysRev.136.B864,PhysRev.140.A1133},
while such relations
have either not been suggested for other properties of the UEG, or, in case
they have been suggested, their validity is still under debate or the exact
form of the relation is unknown. 
For example, several works suggested that the bandnarrowing found in
experiments and calculations of the Alkali
metals~\cite{Mandal2022} can be explained by a bandnarrowing found in
earlier theories of the UEG~\cite{PhysRevLett.62.2718}. However, recent
work~\cite{Haule2022_diagrammatic} finds the bandnarrowing in the UEG
to be much smaller than in the earlier
calculations, which suggests that for this quantity there might not be
a useful relation between the UEG and realistic materials.

On the other hand, it seems plausible that the $Z$ renormalization factor
of quasiparticles at the Fermi surface of a realistic material
might be related to its counterpart in the UEG. 
One reason why $Z$ cannot be included explicitly into standard KS-DFT is that
only one effective potential is used there, which only takes into account
the lattice potential, the Coulomb potential,
and the exchange-correlation
potential.

Recently, we have suggested a moment-functional based spectral-density
functional theory (MFbSDFT)~\cite{momentis},
which computes the spectral function matrix $S_{nm}(E)$ from its
spectral moment matrices
\bege\label{eq_def_specmommat}
M^{(I)}_{nm}=\frac{1}{\hbar}\int d\, E E^I S_{ nm}(E),
\ee
where $I=1,2,3,\dots$.
The key assumption of this approach is that the spectral moment matrices $M^{(I)}_{nm}$
can be computed from the KS-Hamiltonian without correlation, i.e., only with the local
exchange, and additional correction terms $M^{(I+)}_{nm}$, which can be obtained
from moment potentials $\mathcal{V}^{(I+)}(\vn{r})$:
\bege\label{eq_mom_pot_def}
M^{(I+)}_{nm}=\int d^3 r \mathcal{V}^{(I+)}(\vn{r})
\left[\phi_{n}(\vn{r})\right]^{*}
\phi_{m}(\vn{r}),
\ee
where $\phi_{n}(\vn{r})$
are orthonormalized basis functions. We have suggested~\cite{momentis} that these
moment potentials $\mathcal{V}^{(I+)}(\vn{r})$ are likely to be given
by universal functionals of the charge density, similar to the
exchange correlation functional.

In practical calculations one will have to choose the maximal $I$
in Eq.~\eqref{eq_def_specmommat}, i.e., one will choose $I\le 3$, or
$I\le 5$, or $I\le 7$, or $I\le 9$, $\dots$.
By increasing the maximum $I$ one increases the precision of MFbSDFT.
By increasing the maximum $I$ one may also employ more and more properties
of the UEG through the additional moment potentials. For example $I\le 1$
requires $Z=1$, but already with $I\le 3$ it is possible to impose $Z\le1$.
However, the moment potentials $\mathcal{V}^{(I+)}(\vn{r})$
with high $I$ are still unknown. For $I=1$ and $I=2$ they may be obtained
from models of the spectral moments of the UEG~\cite{PhysRevB.69.045113}.

When MFbSDFT is implemented within a second variation
scheme in FLAPW~\cite{PhysRevB.60.10763,second_variation_soi,momentis},
the number $N$ of basis functions used (i.e.\ the number
of $\phi_n$ in Eq.~\eqref{eq_mom_pot_def} and the number of rows and columns of
$\vn{M}^{(I+)}$) can be significantly smaller
than the total number of FLAPW basis functions. 
Since a $PN\times PN$ matrix has to be diagonalized in the MFbSDFT step
-- the computer time requirement of which scales like $\propto (PN)^3$ -- the overall
computational burden
is similar to a standard KS-DFT calculation as long as $PN$ does not
exceed the number of FLAPW basis functions.

In Ref.~\cite{momdis}
we have suggested computing $M^{(3+)}_{nm}$ from the momentum distribution function
$n_k$ of the UEG when the second moment correction $M^{(2+)}_{nm}$
is known. We have shown that
this may improve the spectra in comparison to standard KS-DFT in some cases.
In Ref.~\cite{momdis} we considered $n_k$ only in the vicinity of the Fermi surface.
In the present work we refine the approach of Ref.~\cite{momdis} further,
notably we pay attention to the normalization of $n_k$ and to its integral up to the
Fermi wave number $k_{\rm F}$.
We show that these properties of the UEG can be reproduced when it is modelled within
the 4-pole approximation. Since the quasiparticle renormalization factor $Z$ in the
UEG is related to the discontinuity of $n_k$ at $k_{\rm F}$~\cite{book_mahan},
our 4-pole model also
includes $Z$ by construction. Our 4-pole model may be used to obtain moment potentials
for $I\le 7$.

There are many cases where KS-DFT does not predict the spectral
properties satisfactorily. In fact,
  apart from  the highest occupied KS-eigenvalue in finite
  systems -- which in principle predicts the negative ionization
  energy -- the KS-eigenvalues have no mathematically rigorous
  relation with the experimental spectra within
  KS-theory~\cite{book_martin_2020}.
Notably, the KS-bandgap in insulators deviates often substantially
from experiment~\cite{PhysRevLett.96.226402}.
In strongly correlated materials the upper and lower Hubbard bands
may be missing in the KS-spectrum~\cite{PhysRevLett.93.156402,PhysRevB.72.155106}.
Even in simple metals such as Na and K the bandwidths may differ significantly
from experiment~\cite{Mandal2022}.
To improve all these spectral properties by MFbSDFT might be possible, but this
would require a sufficient number of moment potentials $\mathcal{V}^{(I+)}(\vn{r})$
with sufficient accuracy, which are not available yet.

However, bandgaps
can often be corrected by LDA+$U$, while missing upper or lower Hubbard bands or wrong
bandwidths often cannot be corrected easily by LDA+$U$. Correcting
the bandgap may therefore define a goal that can be achieved by MFbSDFT with a minimum
number of moment potentials. If universal moment potentials can be found that correct
the bandgap in many insulators, it corroborates the key assumptions
of the MFbSDFT approach.
We show in this work that this is indeed the case: Already the moments with $I\le 3$ are
sufficient to correct the bandgap in many insulators. This provides a strong motivation
to develop suitable models also for the higher spectral
moments of the UEG in order to reproduce
additionally the experimental bandwidths and upper and lower Hubbard bands in future
improvements of the MFbSDFT method.

The rest of this paper is structured as follows:
In Sec.~\ref{sec_specmom_lowerspecmom} we explain how higher
spectral moments may be expressed in terms of lower spectral moments
and additional correction terms. Additionally, we introduce the
zero-bandnarrowing approximation of the UEG.
In Sec.~\ref{sec_threepole}
and in Sec.~\ref{sec_fourpole}
we discuss in detail the three-pole and four-pole approximations of the UEG, respectively. 
In Sec.~\ref{sec_npole}
we describe an $(n+1)$-pole approximation of the UEG,
which further improves the description in particular
of $n_k$.
In Sec.~\ref{sec_opti_2p}
we explain how the two-pole approximation can be optimized for weakly and moderately correlated
systems by leaving out the low-energy satellite band.
In Sec.~\ref{sec_results}
we present the results of first-principles MFbSDFT calculations
based on the moment potentials obtained from these models of the UEG.
This paper ends with a summary in Sec.~\ref{sec_summary}.
In the Appendices we
discuss how to obtain $M^{(2+)}$ from the
model developed in Ref.~\cite{PhysRevB.69.045113}
(Sec.~\ref{app_model_secmom})
and how to compute the spectral function from the first 6 spectral moments 
(Sec.~\ref{sec_app_mfbsdft_6mom}).

\section{Theory}
\subsection{Expressing spectral moments in terms of lower spectral moments and correction terms}
\label{sec_specmom_lowerspecmom}
The matrix elements of the spectral moment $\vn{M}^{(1)}$ are given by~\cite{momentis}
\bege
M^{(1)}_{nm}=T_{nm}+V^{\rm H}_{nm}+V^{\rm X}_{nm},
\ee
where
\bege
T_{nm}=\int d^3 r \phi^{*}_{n}(\vn{r})
\left[
  -\frac{\hbar^2}{2m}\Delta + V(\vn{r})
\right]
\phi_{m}(\vn{r})
\ee
comprises the kinetic energy and the
crystal potential $V(\vn{r})$,
$V^{\rm H}_{nm}$ are the matrix elements of the
Hartree potential, and
\bege
V^{\rm X}_{nm}=-\frac{\hbar^2}{m a_{\rm B}^2}\left[
  \frac{3}{2\pi}
  \right]^{\frac{2}{3}}
\int d^3 r \phi^{*}_{n}(\vn{r})\phi_{m}(\vn{r})\frac{1}{r_s(\vn{r})}
\ee
are the matrix elements of the local exchange potential, where
\bege\label{eq_dimlessdenparam}
r_s(\vn{r})=
\frac{1}{a_{\rm B}}
\left(
\frac{3}{4\pi n_{\rm e}(\vn{r})}
\right)^{\frac{1}{3}}
\ee
is the dimensionless density parameter~\cite{book_mahan}, and $a_{\rm B}$
is the Bohr radius. $r_s$ depends on the position $\vn{r}$
through the electron density $n_{\rm e}(\vn{r})$.

We express the second moment matrix $\vn{M}^{(2)}$
in terms of the square of the first moment matrix $\vn{M}^{(1)}$ 
and the correction term $\vn{M}^{(2+)}$~\cite{momentis}:
\bege
\label{eq_secmom_firstmomsq}
\vn{M}^{(2)}=
\left[
  \vn{M}^{(1)}
  \right]^2+\vn{M}^{(2+)}.
\ee
The correction term $\vn{M}^{(2+)}$ may be computed from
the moment potential $\mathcal{V}^{(2+)}$ according to Eq.~\eqref{eq_mom_pot_def}.
A model for $\mathcal{V}^{(2+)}$, which has been derived from the spectral moments of the
UEG, is described in the Appendix~\ref{app_model_secmom}.

In Ref.~\cite{momentis} we suggested using
\bege\label{eq_3rdmom_suggested}
\vn{M}^{(3)}=
\left[
  \vn{M}^{(1)}
  \right]^3+\vn{M}^{(3+)}
\ee
and we have shown that the spectra can be improved in some cases using this expression.
However, Eq.~\eqref{eq_3rdmom_suggested}
is not the only possible form that one may suggest.
For example
\bege\label{eq_3rdmom_alter}
\vn{M}^{(3)}=
\frac{1}{2}
\vn{M}^{(1)}
\vn{M}^{(2)}
+
\frac{1}{2}
\vn{M}^{(2)}
\vn{M}^{(1)}
  +\vn{M}^{(3+)}
\ee
is a priori also a possible form.

When we describe the UEG with the method of spectral moments and
set $M^{(2+)}$ and  $M^{(3+)}$ to $k$-independent constants (in the UEG
the $M^{(I+)}$ are real-valued numbers and not matrices, therefore
we do not use bold-face when $M^{(I+)}$ or $M^{(I)}$ refer to the UEG)
we find that e.g.\ the bandnarrowing may differ depending on whether
Eq.~\eqref{eq_3rdmom_suggested}
or Eq.~\eqref{eq_3rdmom_alter} is used,
because in general
\bege
M^{(1)}
M^{(2)}\ne \left[
 M^{(1)}
  \right]^3.
\ee

For given $M^{(1)}$, $M^{(2)}$ and  $M^{(3)}$ Eq.~\eqref{eq_3rdmom_suggested}
and Eq.~\eqref{eq_3rdmom_alter} can be solved for $M^{(3+)}$.
Therefore, if we used a $k$-dependent $M^{(3+)}$ we could obviously
obtain the same
results from both Eq.~\eqref{eq_3rdmom_suggested}
and Eq.~\eqref{eq_3rdmom_alter} (clearly, the $M^{(3+)}$ to be used together
with Eq.~\eqref{eq_3rdmom_suggested} would differ from
the $M^{(3+)}$ to be used together
with Eq.~\eqref{eq_3rdmom_alter} in order to obtain the same $M^{(3)}$
for given $M^{(1)}$ and $M^{(2)}$).
However, when we choose $M^{(3+)}$ to be independent of $k$ the question
arises of
whether Eq.~\eqref{eq_3rdmom_suggested}
or Eq.~\eqref{eq_3rdmom_alter} is the better alternative.

More generally, we might even consider
\bege\label{eq_3rdmom_alter_general}
\begin{aligned}
\vn{M}^{(3)}=&
\frac{\gamma}{2}
\vn{M}^{(1)}
\vn{M}^{(2)}
+
\frac{\gamma}{2}
\vn{M}^{(2)}
\vn{M}^{(1)}\\
&+(1-\gamma)
\left[
  \vn{M}^{(1)}
  \right]^3
+\vn{M}^{(3+)}
\end{aligned}
\ee
with a parameter $\gamma$ that could be chosen to optimize the results.
For example, one might determine $\gamma$ so that the bandnarrowing of the UEG
is reproduced as well as possible
by the method of spectral moments with $k$-independent $M^{(I+)}$.

At $r_s=4$ the bandnarrowing in the UEG is only 4-7\% according to the
variational diagrammatic Monte Carlo
calculations of Ref.~\cite{Haule2022_diagrammatic}.
However, values of the bandnarrowing for the full range of variation
of $r_s$ (see Table~\ref{tab_rsminmax}) as used in first-principles
calculations have not yet been published for the variational
diagrammatic Monte Carlo method.
For small bandnarrowings that do not exceed 4-7\% it is plausible that a
zero-bandnarrowing approximation of the UEG may yield useful results.
It will become clear in Sec.~\ref{sec_threepole},
Sec.~\ref{sec_fourpole}, and
Sec.~\ref{sec_npole} that it is indeed very instructive and insightfull
to develop and to investigate such a
zero-bandnarrowing approximation of the UEG.

Interestingly, it is rather easy to determine the coefficient $\gamma$ in
Eq.~\eqref{eq_3rdmom_alter_general}
so that the bandnarrowing is precisely zero.
The basic observation is that we may rewrite Eq.~\eqref{eq_secmom_firstmomsq}
as follows:
\bege
\begin{aligned}
\vn{M}^{(2+)}=&
\vn{M}^{(2)}-
\left[
\vn{M}^{(1)}
\right]^2=\\
=&
\frac{1}{\hbar}
\int
\left[
E-\vn{M}^{(1)}
\right]^2
\vn{S}(E)dE.
\end{aligned}
\ee
Generalizing this expression to $I=3$ we obtain
\bege\label{eq_m3+_altermethod}
\begin{aligned}
  &\vn{M}^{(3+)}=
  \frac{1}{\hbar}
{\rm Re}\left\{
\int
\left[
E-\vn{M}^{(1)}
\right]^3
\vn{S}(E)dE
\right\}
=\\
&=\vn{M}^{(3)}+2
\left[
\vn{M}^{(1)}
\right]^3
-\frac{3}{2}\left[
\vn{M}^{(2)}
\vn{M}^{(1)}
+
\vn{M}^{(1)}
\vn{M}^{(2)}\right],
\end{aligned}
\ee
where we define the real-part of a
matrix $\vn{A}$ as
\bege\label{eq_real_part}
    {\rm Re}\vn{A}=\frac{1}{2}
    \left[
      \vn{A}
+
      \vn{A}^{\dagger}
      \right].
\ee
Using Eq.~\eqref{eq_real_part} ensures that the
spectral moment correction is hermitian.
Comparing Eq.~\eqref{eq_m3+_altermethod}
to Eq.~\eqref{eq_3rdmom_alter_general} we
find that they become equivalent when we set $\gamma=3$.
In practice, $\vn{M}^{(2+)}$ and  $\vn{M}^{(3+)}$ are computed from
suitable moment potentials according to Eq.~\eqref{eq_mom_pot_def}.
Next, one computes $\vn{M}^{(2)}$ from Eq.~\eqref{eq_secmom_firstmomsq}.
Finally, one computes $\vn{M}^{(3)}$ from Eq.~\eqref{eq_m3+_altermethod}.
When one uses this recipe to compute the bandstructure of the UEG one finds
numerically that the bandnarrowing is zero. It is likely that this
zero-bandnarrowing property can also be proven analytically, which we leave
for future work.

This recipe produces also a zero bandnarrowing in the UEG when we include more moments.
For $I=4$ we have
\bege\label{eq_momcorr4}
\begin{aligned}
  &\vn{M}^{(4+)}=
  \frac{1}{\hbar}
{\rm Re}\left[
  \int
\left[
E-\vn{M}^{(1)}
\right]^4
\vn{S}(E)dE
\right]
=\vn{M}^{(4)}\\
&-4
{\rm Re}\left[
\vn{M}^{(3)}\vn{M}^{(1)}
\right]
+6
{\rm Re}\left[
\vn{M}^{(2)}\left[
\vn{M}^{(1)}
\right]^2
\right]
-3
\left[
\vn{M}^{(1)}
\right]^4
\end{aligned}
\ee
and for $I=5$ we have
\bege\label{eq_momcorr5}
\begin{aligned}
  &\vn{M}^{(5+)}=
  \frac{1}{\hbar}
  {\rm Re}\left[
\int
\left[
E-\vn{M}^{(1)}
\right]^5
\vn{S}(E)dE\right]=
\vn{M}^{(5)}
\\
&
-10
{\rm Re}\left[
\vn{M}^{(2)}\left[
\vn{M}^{(1)}
\right]^3
\right]
+10
{\rm Re}\left[
\vn{M}^{(3)}\left[
\vn{M}^{(1)}
\right]^2\right]\\
&-5
{\rm Re}\left[
\vn{M}^{(4)}\vn{M}^{(1)}
\right]+4
\left[
\vn{M}^{(1)}
\right]^5.
\end{aligned}
\ee
Hence, the general expression is of the
form
\bege
\vn{M}^{(I)}=\vn{M}^{(I+)}+\sum_{J=1}^{I-1}
\gamma_{J}^{\phantom{J}}{\rm Re}
\left[
[\vn{M}^{(1)}]^{J}\vn{M}^{(I-J)}
  \right],
\ee
which contains $I-1$ coefficients $\gamma_{J}$,
which satisfy $\sum_{J=1}^{I-1} \gamma_{J}=1$.
When these coefficients are determined according to 
the expansion of ${\rm Re}[E-\vn{M}^{(1)}]^{I}\vn{S}(E)$ as in the
examples above, the bandnarrowing is zero.

Interestingly, with these expressions not only the bandnarrowing is zero in the UEG.
Importantly, also the spectral weights $a_{k,j}$ in the $n$-pole approximation 
\bege\label{eq_spec_fun_ueg_npole}
S_{k}(E)=\hbar\sum_{j=1}^n a_{k,j}\delta(E-E_{k,j}),
\ee
of the spectral function of the UEG are $k$-independent
when the expressions above are used to obtain
the spectral moments $M^{(I)}$ from $k$-independent moment corrections $M^{(I+)}$.
In this case the $n$ bands $E_{k,j}$ corresponding to the $n$ poles are simply
parabolas
\bege\label{eq_ueg_parabolas}
E_{k,j}=\frac{\hbar^2 k^2}{2m}+E_{0,j},
\ee
where the Gamma-point energies $E_{0,j}$ of the bands are determined by the
moment corrections $M^{(I+)}$. When the spectral weights $a_{k,j}$ are
$k$-independent in Eq.~\eqref{eq_spec_fun_ueg_npole}, i.e., when
$a_{k,j}=a_{j}$, the spectral function $S_{k}(E)$ depends on $k$ only through
the $k$-dependence of $E_{k,j}$.

In the next sections we discuss the three-pole and four-pole approximations in detail
and show that one may even take the limit of an infinite number of poles.
Thereby, we show that as the number of poles increases one can reproduce more
and more properties of the UEG. In particular, the momentum distribution function,
the second moment, and the charge response are very well reproduced 
when the number of poles is sufficient (Sec.~\ref{sec_npole}). In 
Sec.~\ref{sec_threepole},
Sec.~\ref{sec_fourpole}, and
Sec.~\ref{sec_npole}
we will always use Eq.~\eqref{eq_m3+_altermethod}
and its generalizations to higher $I$, e.g.\
Eq.~\eqref{eq_momcorr4}
and
Eq.~\eqref{eq_momcorr5}, i.e., we will always neglect the bandnarrowing.
The observation that we may reproduce the momentum distribution function,
the second moment, and the charge response
therefore shows that the zero-bandnarrowing approximation suggested in this
section works very well for the UEG.

In future refinements of this approach one might include the effect of the bandnarrowing in the
UEG. In general, the weights $a_{k,j}$ will then become $k$-dependent. Moreover, the
bands will then not simply be mutually shifted parabolas as in Eq.~\eqref{eq_ueg_parabolas}.
This will turn the mathematically very simple recipes to construct the moment potentials
for MFbSDFT
as presented in Sec.~\ref{sec_threepole},
Sec.~\ref{sec_fourpole}, and
Sec.~\ref{sec_npole}
into very complex high-dimensional multivariate optimization problems, because
the necessary integrations 
can only be performed analytically when the weights $a_{k,j}$ are $k$-independent and when
the band dispersions are available in a simple analytical form, such as Eq.~\eqref{eq_ueg_parabolas}.
However, this is only the case when the zero-bandnarrowing approximation is used.

The moment potentials $\mathcal{V}^{(I+)}$
tend to become steeper and steeper in $r_s$ with increasing $I$:
An important contribution to the moment potentials
is~\cite{momentis}
\bege
\mathcal{V}^{(I+)}(\vn{r})=\frac{c^{(I+)}}{[r_s(\vn{r})]^I}+\dots.
\ee
When one evaluates the integral Eq.~\eqref{eq_mom_pot_def} within the FLAPW method
one expresses $\mathcal{V}^{(I+)}(\vn{r})$ on a radial grid inside the MT-spheres.
In the interstitial region one employs a representation of $\mathcal{V}^{(I+)}(\vn{r})$
in reciprocal space, which is obtained from a fast Fourier transform.
A priori one may therefore expect that the convergence of the integral Eq.~\eqref{eq_mom_pot_def}
may be hampered by the steep increase from $[r_s]^{-I}$.
However, there is a simple solution to avoid this potential difficulty:
We compute the moment potentials $\mathcal{V}^{(I+)}(r_s)$ from the UEG as explained
in the following sections. Instead of using them directly in the
integral Eq.~\eqref{eq_mom_pot_def}
we compute the $I$-th root:
\bege
\bar{\mathcal{V}}^{(I+)}(r_s)=
\left[
\mathcal{V}^{(I+)}(r_s)
  \right]^{\frac{1}{I}}.
\ee
Using the $I$-th root moment potentials
we first compute the matrix elements
\bege\label{eq_mom_pot_def_root}
\bar{M}^{(I+)}_{nm}=\int d^3 r \bar{\mathcal{V}}^{(I+)}(\vn{r})
\left[\phi_{n}(\vn{r})\right]^{*}
\phi_{m}(\vn{r}),
\ee
and from them we obtain the
spectral moment corrections:
\bege
\vn{M}^{(I+)}
=
\left[
  \bar{\vn{M}}^{(I+)}
  \right]^I.
\ee

\subsection{The three-pole approximation}
\label{sec_threepole}
In the three-pole approximation of the UEG the spectral function $S_{k}(E)$
is given by
\bege\label{eq_spec_fun_ueg_threepole}
S_{k}(E)=\hbar\sum_{j=1}^3 a_{k,j}\delta(E-E_{k,j}),
\ee
where $a_{k,j}$ and $E_{k,j}$ are the spectral weights and the spectral poles,
respectively.
This model predicts the momentum distribution to be
\bege\label{eq_ueg_nk_model}
n_k=\sum_{j=1}^3 f(E_{k,j})a_{k,j}.
\ee
We assume that
$E_{k,1}<E_{\rm F}$, $E_{k,3}>E_{\rm F}$, and $E_{k_{\rm F},2}=E_{\rm F}$, i.e.,
only $E_{k,2}$ crosses the Fermi level $E_{\rm F}$
at the Fermi wave number $k_{\rm F}$.
Consequently, setting
$a_{k_{\rm F},1}=n_{k_{{\rm F}+}}$,
$a_{k_{\rm F},2}=n_{k_{{\rm F}-}}-n_{k_{{\rm F}+}}$
and $a_{k_{\rm F},3}=1-n_{k_{{\rm F}-}}$ in  Eq.~\eqref{eq_ueg_nk_model}
ensures that we reproduce  $n_{k_{{\rm F}+}}$, $n_{k_{{\rm F}-}}$, and
hence also the step $Z=n_{k_{{\rm F}-}}-n_{k_{{\rm F}+}}$ of $n_k$ at $k_{\rm F}$.
Suitable models for  $n_{k_{{\rm F}-}}$ and $n_{k_{{\rm F}+}}$
are given in Ref.~\cite{PhysRevB.66.235116}.
They can be used to determine the spectral weights $a_{k_{\rm F},j}$
at the Fermi wave number.

Due to the assumption $E_{k_{\rm F},2}=E_{\rm F}$ we can
determine the energy of the second pole at the Fermi wave number by the
Fermi energy.
When we model the UEG within KS-DFT the bandenergy is
given by
\bege
E^{\rm KS}_{k}=\frac{\hbar^2 k^2}{2m}+\frac{d}{dn_{\rm e}}
\left[
n_{\rm e} E^{\rm xc}(n_{\rm e})
\right],  
\ee
where $E^{\rm xc}(n_{\rm e})$ is the exchange-correlation energy per
particle of the
UEG with charge density $n_{\rm e}$.
Consequently, KS-DFT predicts the Fermi energy of the UEG
to be
$E_{\rm F}=E^{\rm KS}_{k_{\rm F}}$,
where
\bege\label{eq_fermi_wave_number}
k_{\rm F}=(\bar{\alpha} r_s a_{\rm B})^{-1}
\ee
is
the Fermi wave number, and
$\bar{\alpha}=[4/(9\pi)]^{1/3}$.
$E_{\rm F}$ as predicted by KS-DFT is consistent with
the theorem of Seitz, which relates the
Fermi energy to the ground state
energy per particle $E_{\rm g}$~\cite{book_mahan}:
\bege\label{eq_seitz}
E_{\rm F}=\frac{d}{dn_{\rm e}}
\left[
n_{\rm e} E_{\rm g}(n_{\rm e})
\right].  
\ee
Therefore, we may expect that KS-DFT predicts the Fermi energy
of the UEG sufficiently accurately.
This provides us with a relation for the energy of the second
pole $E_{k_{\rm F},2}$:
\bege
E_{k_{\rm F},2}=
\frac{\hbar^2}{2 m a_{\rm B}^2}\frac{1}{\left[\bar{\alpha}r_s\right]^2}+
V^{\rm xc},
\ee
where
\bege
V^{\rm xc}=\frac{d}{dn_{\rm e}}\left[
n_{\rm e} E^{\rm xc}(n_{\rm e})
  \right]
\ee
is the exchange-correlation potential.

The retarded Green's function can be obtained easily from the
spectral function~\cite{book_Nolting}:
\bege\label{eq_retarded_green}
\begin{aligned}
  G_{k}(E)&=\int_{-\infty}^{\infty}dE'\frac{S_{k}(E')}{E-E'+i0^{+}}
  \\
  &=\hbar\sum_{j=1}^3\frac{ a_j}{E-E_{k,j}+i0^{+}},
\end{aligned}  
\ee
where we set $a_{k,j}=a_{j}$, which is valid within the zero-bandnarrowing
approximation of the UEG.
Defining the noninteracting retarded Green's function
by
\bege\label{eq_retarded_green_noint}
G_{0,k}(E)=\frac{\hbar}{E-E_{k}^{\rm KS}+i0^{+}},
\ee
we may compute the retarded selfenergy from
\bege
\begin{aligned}
\Sigma_{k}(E)&=\hbar\left[
  G_{0,k}^{-1}-G_{k}^{-1}
  \right]\\
&=E-E^{\rm KS}_{k}+i0^{+}
-\frac{1}
{\sum\limits_{j=1,3}\frac{a_j}{E-E_{k,j}+i0^{+}}}.
\end{aligned}
\ee
We find
\bege
\left.
\frac{\partial \Sigma_{k_{\rm F}}}{\partial E}
\right|_{E=E_{\rm F}}=
1-\frac{1}{a_2}.
\ee
Consequently, the renormalization coefficient
at $E=E_{\rm F}$ is~\cite{book_mahan}
\bege\label{eq_Z_renor_facto}
Z_{\rm F}=\frac{1}{1-\left.
\frac{\partial \Sigma_{k_{\rm F}}}{\partial E}
\right|_{E=E_{\rm F}}}=a_2=n_{k_{\rm F-}}-n_{k_{\rm F+}}.
\ee
The Green function of the three-pole approximation
yields therefore a renormalization coefficient that
is consistent with the jump of the momentum distribution function.
This corroborates the overall consistency of our three-pole
spectral function.

Many applications of the method of spectral
moments~\cite{book_Nolting,nickel_PhysRevB.40.5015,nickel_Borgiel1990}
do not
consider a finite imaginary part of the self-energy. Similarly,
in Eq.~\eqref{eq_retarded_green} and in
Eq.~\eqref{eq_retarded_green_noint}
we
use $i0^{+}$ (where $0^{+}$ is a positive infinitesimal)
only to ensure the proper analytical behaviour of the Green's function.
How finite imaginary parts of the self-energy may arise
within the method of spectral moments
is a very interesting question. Ref.~\cite{Nolting_1980} has already
suggested using Gaussians instead of delta-functions in the
$n$-pole approximation as a possible way.
Before turning back to the discussion of the three-pole approximation
we describe in the following an alternative perspective on the
question of finite imaginary parts of the self-energy within
the method of spectral moments, because it extends the discussion
of Eq.~\eqref{eq_retarded_green} through Eq.~\eqref{eq_Z_renor_facto}
to the case $n\rightarrow \infty$, where $n$ is the number of poles
in the $n$-pole model, which will be discussed again in
Sec.~\ref{sec_npole}.

The generalizations of
Eq.~\eqref{eq_spec_fun_ueg_threepole}
and Eq.~\eqref{eq_retarded_green}
to $n$ poles are
\bege\label{eq_spec_fun_ueg_npole123}
S_{k}(E)=\hbar\sum_{j=1}^n a_{k,j}\delta(E-E_{k,j}),
\ee
and
\bege\label{eq_retarded_green_npoles}
  G_{k}(E)=\hbar\sum_{j=1}^n\frac{ a_j}{E-E_{k,j}+i0^{+}}.
\ee
In the limit of $n\rightarrow \infty$ the
discrete spectral weights in Eq.~\eqref{eq_spec_fun_ueg_npole123}
turn into a continuous distribution function:
\bege\label{eq_spec_fun_ueg_conti}
S_{k}(E)=\int_{-\infty}^{\infty}dE' \delta(E-E')S_{k}(E'),
\ee
i.e.,
\bege
\hbar a_{k,j}\rightarrow S_{k}(E')
\ee
and
\bege
\sum_{j=1}^n\rightarrow \int_{-\infty}^{\infty}dE'
\ee
are the transformations to be performed on Eq.~\eqref{eq_spec_fun_ueg_npole}
in order to transform it from its discrete form into
the continuous form of Eq.~\eqref{eq_spec_fun_ueg_conti}.
The continuous form of Eq.~\eqref{eq_retarded_green_npoles}
is given by the first line of Eq.~\eqref{eq_retarded_green}.
Thus, in the limit $n\rightarrow \infty$ the $n$-pole approximation
can describe any given spectral function, because Eq.~\eqref{eq_spec_fun_ueg_conti}
is a trivial identity that holds for any given spectral function,
and the first line of Eq.~\eqref{eq_retarded_green} is generally valid as well.

When a retarded Green's function is given in the form
\bege\label{eq_green_with_self}
G_{k}(E)=\frac{1}{E-E_{k}-\Sigma_{k}(E)},
\ee
the spectral function can be obtained from~\cite{book_Nolting}
\bege\label{eq_spec_func_imagpart}
\begin{aligned}
  S_{k}(E)&=-\frac{1}{\pi}{\rm Im}\left[
G_{k}(E)
\right]=\\
  &=-\frac{1}{\pi}
  \frac{{\rm Im}\left[
\Sigma_{k}(E)
\right]}{
    \left(E-E_{k}-
{\rm Re}\left[
\Sigma_{k}(E)
\right]
    \right)^2
    +
    \left(
{\rm Im}\left[
\Sigma_{k}(E)
\right]
    \right)^2
  }.
\end{aligned}
\ee
Therefore, in the limit $n\rightarrow \infty$ the
$n$-pole approximation becomes exact and it can
accommodate the finite imaginary part of the self-energy, because
the Green's function in Eq.~\eqref{eq_green_with_self} can be
recovered from the first line in Eq.~\eqref{eq_retarded_green}
when Eq.~\eqref{eq_spec_func_imagpart} is inserted into it.

At the Fermi energy the imaginary part of the self-energy is often
zero. In those cases one can model the spectral function by~\cite{book_mahan}
\bege\label{eq_model_specfun_cont_and_disc}
S_{k}(E)=\bar{S}_{k}(E)+a^{\phantom{F_{F}}}_{{\rm F}k}\delta(E-E_{\rm F}),
\ee
where $\bar{S}_{k}(E)$ is a smooth function of $E$.
For $E\ne E_{\rm F}$ $S_{k}(E)=\bar{S}_{k}(E)$
can be obtained easily from Eq.~\eqref{eq_spec_func_imagpart}
when ${\rm Im}\left[
\Sigma_{k}(E)
\right]<0$. In order to extract the coefficient $a^{\phantom{F_{F}}}_{{\rm F}k}$
in Eq.~\eqref{eq_model_specfun_cont_and_disc} at the Fermi
wave number $k_{\rm F}$
one may
substitute
\bege
{\rm Im}\left[
\Sigma_{k_{\rm F}}(E_{\rm F})
\right]\rightarrow
-\Gamma
\ee
when ${\rm Im}\left[
\Sigma_{k_{\rm F}}(E_{\rm F})
\right]=0$
and take the limit $\Gamma\rightarrow 0$:
\bege
\begin{aligned}
  &a^{\phantom{F_{F}}}_{{\rm F},k_{\rm F}}=
  -\frac{1}{\pi}
  \lim_{\eta\rightarrow 0}
  \lim_{\Gamma\rightarrow 0}
  \int_{E_{\rm F}-\eta}^{E_{\rm F}+\eta}
      {\rm Im} \left[G_{k_{\rm F}}(E)\right] dE=\\
&=
  \lim_{\eta\rightarrow 0}
  \lim_{\Gamma\rightarrow 0}
  \int_{E_{\rm F}-\eta}^{E_{\rm F}+\eta}
  \frac{\frac{\Gamma}{\pi}  dE  }{
    \left(E-E_{k}-
{\rm Re}\left[
\Sigma_{k_{\rm F}}(E)
\right]
    \right)^2
    +
    \Gamma^2
  }\\
&=
  \lim_{\eta\rightarrow 0}
  \int_{E_{\rm F}-\eta}^{E_{\rm F}+\eta}
\delta\left(E-E_{k_{\rm F}}-
{\rm Re}\left[
\Sigma_{k_{\rm F}}(E)
\right]
    \right)dE
    \\
    &=\frac{1}{1-
      \left.
      \frac{\partial {\rm Re}\Sigma_{k_{\rm F}}(E)}{\partial E}
    \right|_{E=E_{\rm F}}},    
\end{aligned}
\ee
where $E_{\rm F}=E_{k_{\rm F}}+
{\rm Re}\left[
\Sigma_{k_{\rm F}}(E_{\rm F})
\right]$. This result corresponds to
Eq.~\eqref{eq_Z_renor_facto}
in the limit $n\rightarrow \infty$.

Now we return to the discussion of the three-pole model.
In order to determine the remaining two poles $E_{k_{\rm F},1}$
and $E_{k_{\rm F},3}$
we may employ the models for the spectral moments $M_{k}^{(1)}$
and $M_{k}^{(2)}$ that have been developed in Ref.~\cite{PhysRevB.69.045113}.
The first moment is
\bege
M_{k_{\rm F}}^{(1)}=\frac{\hbar^2k_{\rm F}^2 }{2m}-\frac{\hbar^2}{m a_{\rm B}^2}
\left[
\frac{3}{2\pi}
  \right]^{\frac{2}{3}}\frac{1}{r_s}
\ee
and
the expressions for $M_{k_{\rm F}}^{(2)}$ are given in Appendix~\ref{app_model_secmom}.
This provides us with the two equations
\bege
M_{k_{\rm F}}^{(1)}=a_{k_{\rm F},1}E_{k_{\rm F},1}+a_{k_{\rm F},2}E_{k_{\rm F},2}+a_{k_{\rm F},3}E_{k_{\rm F},3}
\ee
and
\bege
M_{k_{\rm F}}^{(2)}=a_{k_{\rm F},1}E^2_{k_{\rm F},1}+a_{k_{\rm F},2}E^2_{k_{\rm F},2}+a_{k_{\rm F},3}E^2_{k_{\rm F},3}
\ee
for the two yet unknown poles $E_{k_{\rm F},1}$ and $E_{k_{\rm F},3}$.
In general these equations have two solutions. However, due to the assumption
$E_{k,1}<E_{k,2}<E_{k,3}$ of our model
we need to consider only the solution
\bege
E_{k_{\rm F},1}=\frac{-B-\sqrt{B^2-4AC}}{2 A},
\ee
where
\bege
A=a_{k_{\rm F},1}a_{k_{\rm F},3}+a_{k_{\rm F},1}^2,
\ee
\bege
B=-2a_{k_{\rm F},1}\left[
M_{k_{\rm F}}^{(1)}-E^{\rm KS}_{k_{\rm F}}a_{k_{\rm F},2}
  \right],
\ee
and
\bege
\begin{aligned}
C&=\left[
M_{k_{\rm F}}^{(1)}-E^{\rm KS}_{k_{\rm F}}a_{k_{\rm F},2}
\right]^2
-M_{k_{\rm F}}^{(2)}a_{k_{\rm F},3}\\
&+\left[
  E^{\rm KS}_{k_{\rm F}}
  \right]^2a_{k_{\rm F},3}a_{k_{\rm F},2}.
\end{aligned}
\ee
Finally, the energy of the third
pole may be computed from
\bege
E_{k_{\rm F},3}=\frac{M_{k_{\rm F}}^{(1)}-a_{k_{\rm F},1}E_{k_{\rm F},1}-a_{k_{\rm F},2}E_{k_{\rm F},2}}{a_{k_{\rm F},3}}
\ee
and
the spectral moments $M_{k_{\rm F}}^{(I)}$ with $I=3,4,5$ may be obtained from
\bege
M_{k_{\rm F}}^{(I)}=a_{k_{\rm F},1}E^I_{k_{\rm F},1}
+a_{k_{\rm F},2}E^I_{k_{\rm F},2}
+a_{k_{\rm F},3}E^I_{k_{\rm F},3}.
\ee

From these results one may extract the moment potentials
$\mathcal{V}^{(3+)}$,
$\mathcal{V}^{(4+)}$,
and
$\mathcal{V}^{(5+)}$ as explained in the preceding section.
Using them, one may perform MFbSDFT calculations using the first
6 spectral moments.
In Ref.~\cite{momentis}
we have already explained in detail how MFbSDFT calculations are performed
based on the first 4 spectral moment matrices.
The only major change when using the first 6 spectral moments is the
construction of the spectral function, which we describe in detail
in the Appendix~\ref{sec_app_mfbsdft_6mom}.
While this approach improves the spectra in some cases,
it also has a severe shortcoming:
The momentum distribution function should be
normalized~\cite{PhysRevB.50.1391},
i.e.,
\bege
\int
dk
k^2
\left[
  n_{k}-
\theta(k_{\rm F}-k)
  \right]=0
\ee
should be satisfied, which is not the case,
because Eq.~\eqref{eq_ueg_nk_model} is so constructed
that $n_k$ is reproduced at the Fermi surface, while
no use is made of this normalization constraint.

In the following we will refer to the bands
$E_{k,j}<E_{k}^{\rm KS}$ as satellite bands.
In the three-pole approximation discussed in this
section $E_{k,1}$ is a satellite band, while $E_{k,2}$ could be
called the KS-band.
The motivation for the name satellite band
comes from the observation of a low-energy
valence-band satellite peak in the spectrum of
Ni, which can be reproduced with the method of
spectral moments~\cite{nickel_PhysRevB.40.5015,nickel_Borgiel1990}.
While a quantitative relation between the satellite bands
in the $n$-pole approximation of the UEG and the satellite peaks in the
photoemission spectra of several real materials has not yet been established,
it seems at least plausible that the satellite bands
in the $n$-pole approximation of the UEG may be considered as 
precursors of the satellite peaks in the spectra of real materials.
The problem of the three-pole approximation proposed above is that
it may give too much relative weight to the satellite bands.

We define an averaged momentum distribution by 
\bege\label{eq_n_average<}
N_{<}=\frac{3}{k_{\rm F}^3}\int_{0}^{k_{\rm F}}k^2 n_{k}d k,
\ee
which considers only $k<k_{\rm F}$ in the average.
A similar averaged momentum distribution can be defined
for $k>k_{\rm F}$:
\bege
\label{eq_n_average>}
N_{>}=\frac{3}{k_{\rm F}^3}\int_{k_{\rm F}}^{\infty}k^2 n_{k} d k.
\ee
Due to the normalization of $n_k$ we have
\bege
N_{<}+N_{>}=1.
\ee

While we can satisfy the constraint
of Eq.~\eqref{eq_n_average<} by using  the modified weights
$a_1=N_{<}-Z$, and $a_2=Z$,
and the constraint
of Eq.~\eqref{eq_n_average>} by choosing  $E_{0,1}$ appropriately,
this choice still attributes too much relative weight to the satellite band,
unless the band $E_{k,3}$ cuts the Fermi level.
However, if we so specify a $k$ at which $E_{k,3}$ cuts the Fermi level
that the satellite band has the appropriate weight, 
we have determined all three energies $E_{k,j}$ without making use of the
first moment $M^{(1)}_{k_{\rm F}}$. Since the MFbSDFT approach suggested
in Ref.~\cite{momentis} assumes that the correct first moment is used in the
construction of the moment potentials, not using the right $M^{(1)}_{k_{\rm F}}$ is not
expected to work. In the next section we show that the problem of normalization
of $n_k$ can be solved within the 4-pole approximation.

\subsection{The four-pole approximation}
\label{sec_fourpole}
In order to determine the energies of the 4 poles
we start by setting
\bege
E_2=E_{\rm KS}
\ee
and
\bege
a_2=Z=n_{k_{{\rm F}-}}-n_{k_{{\rm F}+}}
\ee
like in the three-pole model.
Here, to simplify the notation, we
define $E_{\rm KS}=E^{\rm KS}_{k_{\rm F}}=E_{\rm F}$
and $E_{i}=E_{k_{\rm F},i}$.
In order to find $E_1$ we assume that the corresponding
band cuts the Fermi energy at $k_{\rm F1}$
and has the weight
\bege
a_1=n_{k_{{\rm F}+}}.
\ee
Therefore, we solve
\bege
3a_1
\int_{k_{\rm F}}^{k_{\rm F1}}k^2 dk=
a_1
(
k_{\rm F1}^3
-
k_{\rm F}^3
)
=
k_{\rm F}^3N_{>}
\ee
for $k_{\rm F1}$ and
use it to compute $E_1$
according to
\bege
E_1=E_{\rm KS}+\frac{\hbar^2\left[k_{\rm F}^2-k_{\rm F1}^2\right]}{2m}.
\ee
The final solution is
\bege
E_1=E_{\rm KS}+\frac{\hbar^2k_{\rm F}^2
  \left[1-
  \left(
1+\frac{1-N_{<}}{a_1}
    \right)^{\frac{2}{3}}\right]}{2m}.
\ee
Using $E_1$ ensures that the four-pole model $n_k^{(4p)}$ of $n_k$
satisfies Eq.~\eqref{eq_n_average>}.

Next, we determine $E_3$  assuming that the corresponding
band cuts the Fermi energy at $k_{\rm F3}$ and
has the weight
\bege
a_3=n_{0}-n_{k_{{\rm F}-}}.
\ee
Consequently, we require that
\bege
\begin{aligned}
3(a_1+a_2)\int_{0}^{k_{\rm F}}k^2dk+3a_3\int_{0}^{k_{\rm F3}}k^2dk=k_{\rm F}^3N_{<}
  \end{aligned}
\ee
be satisfied. We use this equation to determine $k_{\rm F3}$,
which is given by
\bege
k_{\rm F3}=k_{\rm F}
\left[
\frac{N_{<}-a_1-a_2}{a_3}
  \right]^{\frac{1}{3}}.
\ee
Thus, we find
\bege
E_3=E_{\rm KS}+\frac{\hbar^2k_{\rm F}^2
  \left[
    1-
    \left(
  \frac{N_{<}-a_1-a_2}{a_3}
    \right)^{\frac{2}{3}}\right]}{2m}.
\ee
Employing $E_3$ ensures that the four-pole model $n_k^{(4p)}$ of $n_k$
satisfies Eq.~\eqref{eq_n_average<}.

Finally,
we compute $E_4$ from
\bege
E_4=\frac{M^{(1)}_{k_{\rm F}}-a_1 E_1-a_2 E_2-a_3 E_3}{a_4}
\ee
using
\bege
a_4=1-a_1-a_2-a_3=1-n_{0}.
\ee

\begin{figure}
\includegraphics[angle=-90,width=\linewidth]{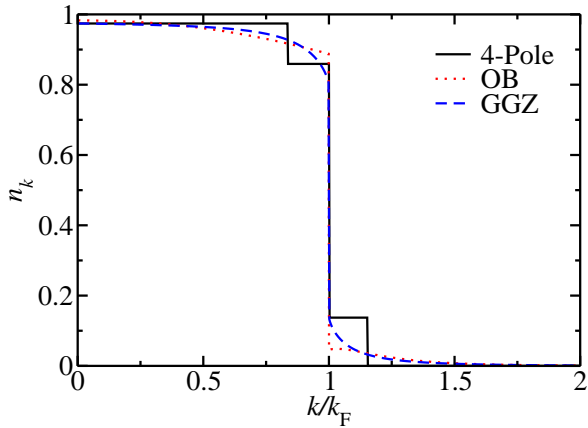}
\caption{\label{fig_4pole_nk}
  Comparison of the momentum distribution function obtained
  within the four-pole model (4-Pole) to the one given in
  Ref.~\cite{PhysRevB.50.1391,PhysRevB.56.9970} (OB)
  and to the one given in
  Ref.~\cite{PhysRevB.66.235116} (GGZ).
  The dimensionless density-parameter is set to $r_s=3$.
}
\end{figure}

In the calculations we use the parametrizations of $n_0$ and
$n_{k_{{\rm F}+}}$ given in Ref.~\cite{PhysRevB.66.235116}.
Due to the differences in the quasiparticle renormalization $Z$
as obtained from  Ref.~\cite{PhysRevB.66.235116} and
Ref.~\cite{Haule2022_diagrammatic}
(see the discussion at the beginning of
Sec.~\ref{sec_results} and Fig.~\ref{fig_zfactor_compare})
we take $Z$ from Ref.~\cite{Haule2022_diagrammatic} and
use it to compute $n_{k_{{\rm F}-}}=n_{k_{{\rm F}+}}+Z$.
In Fig.~\ref{fig_4pole_nk} we plot the momentum distribution
function obtained within the four-pole approximation and
compare it to the models of Ref.~\cite{PhysRevB.66.235116}
and Ref.~\cite{PhysRevB.50.1391,PhysRevB.56.9970}.
In the four-pole model $n_k$ can change only at discrete points $k$,
which is why  $n_k^{(4p)}$ exhibits three jumps.
Only the major jump at $k_{\rm F}$ is also present in the models
of Ref.~\cite{PhysRevB.66.235116} and Ref.~\cite{PhysRevB.50.1391,PhysRevB.56.9970},
while $n_k$ changes smoothly otherwise in the latter models.
Since one may expect that the correct description of the Fermi surface
is particularly important, the two additional jumps in $n_k^{(4p)}$
are not expected to introduce major errors. However, these additional
jumps in $n_k^{(4p)}$ are required to ensure the proper normalization of $n_k^{(4p)}$.
Clearly, $n_k^{(4p)}$ is a significant improvement over standard KS-DFT, which
uses $n_k=\theta(k_{\rm F}-k)$.

In order to test additional properties of our 4-pole model
we start with observing that we have not used the model of $\mathcal{V}^{(2+)}$
discussed in Appendix~\ref{app_model_secmom} in its construction.
Consequently, we may compute $\mathcal{V}^{(2+)}$ from our 4-pole model
and compare it with the result obtained from the expressions
in Appendix~\ref{app_model_secmom}.
In Fig.~\ref{fourpole_compare_secmom} we compare $\mathcal{V}^{(2+)}$
as obtained from the 4-pole model to $\mathcal{V}^{(2+)}$ as given
by the expressions in Appendix~\ref{app_model_secmom}.
The agreement is surprisingly good in view of the  independence of these two
models.
Note that the expressions in Appendix~\ref{app_model_secmom}
are not exact but use the single Slater determinant approximation for one
of the higher-order correlation functions. Moreover, different parametrizations
of the structure factor are available in the literature and the model
of Appendix~\ref{app_model_secmom} yields different results for different parametrizations.
In Sec.~\ref{sec_npole}
we introduce an $n+1$-pole model. We show its results for $\mathcal{V}^{(2+)}$
in Fig.~\ref{fourpole_compare_secmom} as well.
At $r_s=1$ it agrees with the result from the  expressions in Appendix~\ref{app_model_secmom}
quite well, while at $r_s=2$, $r_s=3$, and $r_s=4$ it is instead closer to the 4-pole model.

\begin{figure}
\includegraphics[angle=0,width=\linewidth]{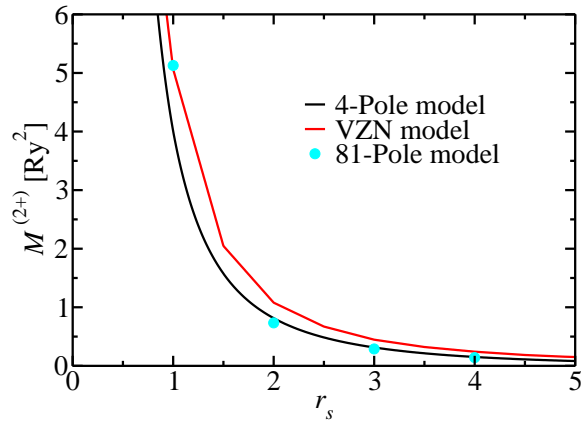}
\caption{\label{fourpole_compare_secmom}
  Comparison between
  $M^{(2+)}_{k_{\rm F}}=\mathcal{V}^{(2+)}$
  as obtained from the 4-pole model,
  the $(n+1)$-pole model of Sec.~\ref{sec_npole}
  with 81-poles,
  and
  from the model of Ref.~\cite{PhysRevB.69.045113}
  discussed in Appendix~\ref{app_model_secmom} (VZN).
  The vertical axis employs the unit of Ry$^{2}$,
  where Ry=$\frac{\hbar^2}{2 m a_{\rm B}^2}$=13.6~eV.
}
\end{figure}

Another property of the UEG that can be used to test our 4-pole model
is the charge response $P(\vn{q},\hbar\omega)$, which is related to the
dielectric function $\epsilon(\vn{q},\hbar\omega)$
by~\cite{book_mahan}
\bege
\epsilon(\vn{q},\hbar\omega)=1-v_{q}P(\vn{q},\hbar\omega).
\ee
Within RPA $\epsilon(\vn{q},\hbar\omega)$
is given by the Lindhard function and the corresponding charge response is
\bege
\bar{P}(\vn{q},E)=\frac{1}{V}\sum_{\vn{k}\sigma}
\frac{f(E_{\vn{k}\sigma})-f(E_{\vn{k}+\vn{q}\sigma})}{E_{\vn{k}\sigma}-E_{\vn{k}+\vn{q}\sigma}+E+i0^{+}}.
\ee

When we model the UEG within KS-DFT
we obtain
\bege
\bar{P}_{0}= \bar{P}(0,0)=-D(E_{\rm F}),
\ee
where 
\bege\label{eq_dos_ueg_ef}
D(E_{\rm F})=\frac{4ma_{\rm B}^2}{(2\pi\hbar)^2\bar{\alpha}r_s}
\ee
is the density of states (DOS) at the Fermi energy.
However, in KS-DFT the induced charge $\delta n_{\rm e}$
changes
the exchange-correlation potential, which in
turn affects the induced charge~\cite{giuliani_vignale_2005}.
Therefore, the induced charge is related to the single-particle charge
response $\bar{P}_{0}$
by
\bege\label{eq_mb_enhancement}
\delta n_{\rm e}=\bar{P}_{0}
\left[
\delta\phi+\frac{\partial V^{\rm xc}}{\partial n_{\rm e}}\delta n_{\rm e}
\right],
\ee
i.e.,
the external perturbing field $\delta \phi$
needs to be combined with $(\partial V^{\rm xc}/\partial n_{\rm e})\delta n_{\rm e}$
into the effective perturbation field
\bege
\delta\phi^{\rm eff}=\delta\phi+
\frac{\partial V^{\rm xc}}{\partial n_{\rm e}}
\delta n_{\rm e}.
\ee
Consequently, the
full $P_{0}=P(0,0)$ is
given by
\bege\label{eq_charge_response_full}
P_{0}=\frac{\delta n_{\rm e}}{\delta\phi}=\frac{\bar{P}_{0}}{1-\bar{P_{0}}
\frac{\partial V^{\rm xc}}{\partial n_{\rm e}}
}.
\ee

\begin{figure}
\includegraphics[angle=0,width=\linewidth]{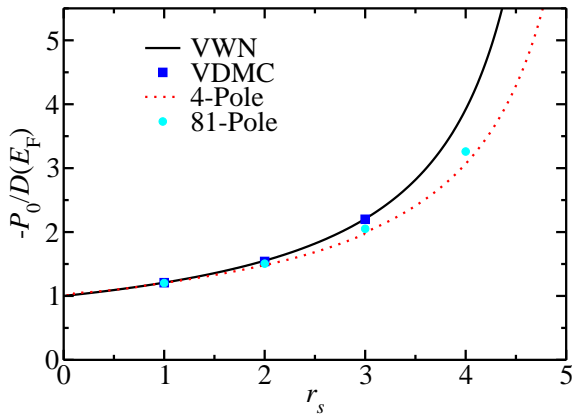}
\caption{\label{fig_charge_polari}
  Ratio of the charge response $P_{0}$ and the
  density of states of the corresponding KS-system
  at the Fermi energy $D(E_{\rm F})$ (Eq.~\eqref{eq_dos_ueg_ef}).
  Four methods are compared. VWN: Eq.~\eqref{eq_charge_response_full}
  using the parametrization of $V^{\rm xc}$ given in Ref.~\cite{vwn}.
  VDMC: Results from variational diagrammatic Monte Carlo
  as given in Ref.~\cite{vdmc_Chen2019}.
  4-Pole: Results for  the 4-pole model as computed
  from Eq.~\eqref{eq_charge_response_full_4p}.
  81-Pole: Results for  the $(n+1)$-pole model as computed
  from Eq.~\eqref{eq_charge_response_full_np1}.
}
\end{figure}

In Fig.~\ref{fig_charge_polari} we show the charge response $P_{0}$ as a function of
the dimensionless density-parameter $r_s$
as computed from Eq.~\eqref{eq_charge_response_full}
using the parametrization of $V^{\rm xc}$ given in Ref.~\cite{vwn}.
Slightly above $r_s=5.2$ there is a well-known charge instability, which is
why $-P_{0}/D(E_{\rm F})$
starts to increase rapidly for $r_s>4$. Recent calculations
of $P_{0}$
by variational diagrammatic Monte Carlo~\cite{vdmc_Chen2019}
are in excellent agreement (squares in the Figure).

In order to compute $P_{0}$ in the 4-pole approximation of the
UEG we need to consider all bands that cross the Fermi energy.
All bands feel a different effective potential $V^{\rm xc}_{i}$.
Therefore, we need to modify Eq.~\eqref{eq_mb_enhancement}
as follows:
\bege
\delta n_{\rm e}=
\sum_{i=1}^{N_{\rm max}}
\bar{P}_{0,i}
\left[
  \delta\phi+
  \frac{\partial E_{0,j}}
{\partial n_{\rm e}}
  \delta n_{\rm e}
\right]
  ,
\ee
where $N_{\rm max}$ is the index of the highest band cutting the Fermi energy
($N_{\rm max}=3$ in the derivations above),
$\bar{P}_{0,1}=D(E_{\rm F})a_{1}k_{\rm F1}/k_{\rm F}$,
$\bar{P}_{0,2}=D(E_{\rm F})a_{2}$,
$\bar{P}_{0,3}=D(E_{\rm F})a_{3}k_{\rm F3}/k_{\rm F}$,
and $E_{0,j}$ is the Gamma-point energy defined in Eq.~\eqref{eq_ueg_parabolas}.
Consequently, we obtain in the 4-pole approximation
\bege\label{eq_charge_response_full_4p}
P^{(4\rm p)}_{0}=\frac{\delta n_{\rm e}}{\delta\phi}=
\frac{
\sum_{i=1}^{N_{\rm max}}\bar{P}_{0,i}
}{1-\sum_{i=1}^{N_{\rm max}}\bar{P}_{0,i}
\frac{\partial E_{0,j}}
{\partial n_{\rm e}}
}.
\ee
Fig.~\ref{fig_charge_polari} shows that the charge response $P_{0}^{(4\rm p)}$
obtained within our 4-pole model of the UEG is in reasonable agreement
with the variational diagrammatic Monte Carlo results.

In Sec.~\ref{sec_npole} we will describe how to increase the number of poles
further. However, the results obtained with 81 poles improve the charge
response only slightly (circles in Fig.~\ref{fig_charge_polari}).
The remaining discrepancies between the $n$-pole approximation and the VDMC
results might result from the zero-bandnarrowing approximation used in this
section. Another possible explanation is
that the $n$-pole approximation differs at least partly from the VDMC results
because the exact $n_k$ is not known and calculations as well as models of
$n_k$ differ in the literature (see e.g.\ Fig.~\ref{fig_zfactor_compare}).

Overall, our four-pole model reproduces the momentum
distribution $n_k$ (Fig.~\ref{fig_4pole_nk}),
the charge response function $P_{0}$ (Fig.~\ref{fig_charge_polari}),
and the second moment (Fig.~\ref{fourpole_compare_secmom}) quite well.
In contrast, we can reproduce only $P_{0}$ within the KS-DFT model of the UEG.

\subsection{The $n$-pole model and the limit $n\rightarrow\infty$}
\label{sec_npole}
The agreement between the properties of the UEG and those predicted
by the 4-pole model can be improved further by considering the following
$(n+1)$-pole model ($n$ is an even number).
We assume that band $n/2$ is the KS-band, i.e.,
\bege
E_{\frac{n}{2}}=E_{\rm KS}
\ee
and
\bege
a_{\frac{n}{2}}=Z.
\ee
The bands $n/2+1,\dots,n$ are higher in energy than the KS-band, but they are
all assumed to cut the Fermi energy.
The bands $1,\dots, n/2-1$ are lower in energy than the KS-band and therefore they
cut the Fermi level as well.

We divide the range from $k=0$ to $k=k_{\rm F}$ into
$n/2$ pieces, where the $i$-th piece is defined by the lower boundary
\bege
k_{{\rm min},i}=k_{\rm F}\frac{1-\delta}{\frac{n}{2}}
\left(
n-i
\right)
\ee
and the upper boundary
\bege
k_{{\rm max},i}=k_{\rm F}\frac{1-\delta}{\frac{n}{2}}(n+1-i),
\ee
where $i=\frac{n}{2}+1,\dots,n$,
and the positive infinitesimal $\delta$ ensures that
$k_{{\rm max},\frac{n}{2}+1}<k_{\rm F}$.

Typically, $n_{(2k_{\rm F})}\ll n_{k_{\rm F}}$.
Consequently, we consider the interval
from $k=k_{\rm F}$ up to $k=2k_{\rm F}$
and divide it into $n/2-1$ pieces,
where the $i$-th piece is defined by the lower boundary
\bege
k_{{\rm min},i}=k_{\rm F}
\left[
  1+\delta+\frac{\frac{n}{2}-1-i}{\frac{n}{2}-1}
\right]
\ee
and the upper boundary
\bege
k_{{\rm max},i}=k_{\rm F}
\left[
  1+\delta+\frac{\frac{n}{2}-i}{\frac{n}{2}-1}
\right].
\ee
Here, $i=1,\dots,\frac{n}{2}-1$,
and the positive infinitesimal $\delta$ ensures that
$k_{{\rm min},\frac{n}{2}-1}>k_{\rm F}$.

We define the corresponding weights as
\bege
a_{i}=n_{k_{{\rm min},i}}-n_{k_{{\rm max},i}}
\ee
for $i=2,\dots \frac{n}{2}-1$
and for $i=\frac{n}{2}+1,\dots, n$.
These weights are always positive if the derivative
\bege
\frac{d n_k}{d k}<0
\ee
is always negative. This is the case for the
parametrization of $n_{k}$ given in Ref.~\cite{PhysRevB.66.235116}.
The weight of the first band we set to
\bege
a_{1}=n_{0}-\sum_{i=2}^n a_{n}.
\ee

The corresponding energies at $k_{\rm F}$
are given by
\bege
E_{i}=E_{\rm KS}+\frac{\hbar^2}{2m}
\left[
k_{\rm F}^2-\frac{(k_{{\rm min},i}+k_{{\rm max},i})^2}{4}
  \right]
\ee
for $i=1,\dots \frac{n}{2}-1$
and for $i=\frac{n}{2}+1,\dots, n$.
Finally, we assume that the band $n+1$ is highest in energy
and does not cut the Fermi energy.
We set its weight to
\bege
a_{n+1}=1-n_{0}
\ee
and determine
its energy from the first moment:
\bege
E_{n+1}=\frac{1}{a_{n+1}}
\left[
  M^{(1)}_{k_{\rm F}}
  -\sum_{i=1}^n
  a_n E_n
  \right].
\ee

The moments $M^{(I)}_{k_{\rm F}}$ for $I>1$ can be computed from
\bege
M^{(I)}_{k_{\rm F}}=\sum_{j=1}^{n+1}a_{j}E_{j}^{I},
\ee
while
the charge response may be obtained from
\bege
\label{eq_charge_response_full_np1}
P_{0}^{([n+1]{\rm p})}=\frac{\delta n_{\rm e}}{\delta \phi}=
\frac{
\mathcal{G}
}
{
  1-\frac{\partial V^{\rm xc}}{\partial n_{\rm e}}\mathcal{G}  
  },
\ee
where
\bege
\mathcal{G}=D(E_{\rm F})\sum_{i=1}^{n}a_i \frac{k_{{\rm min},i}+k_{{\rm max},i}}{2 k_{\rm F}}.
\ee

\begin{figure}
\includegraphics[angle=0,width=\linewidth]{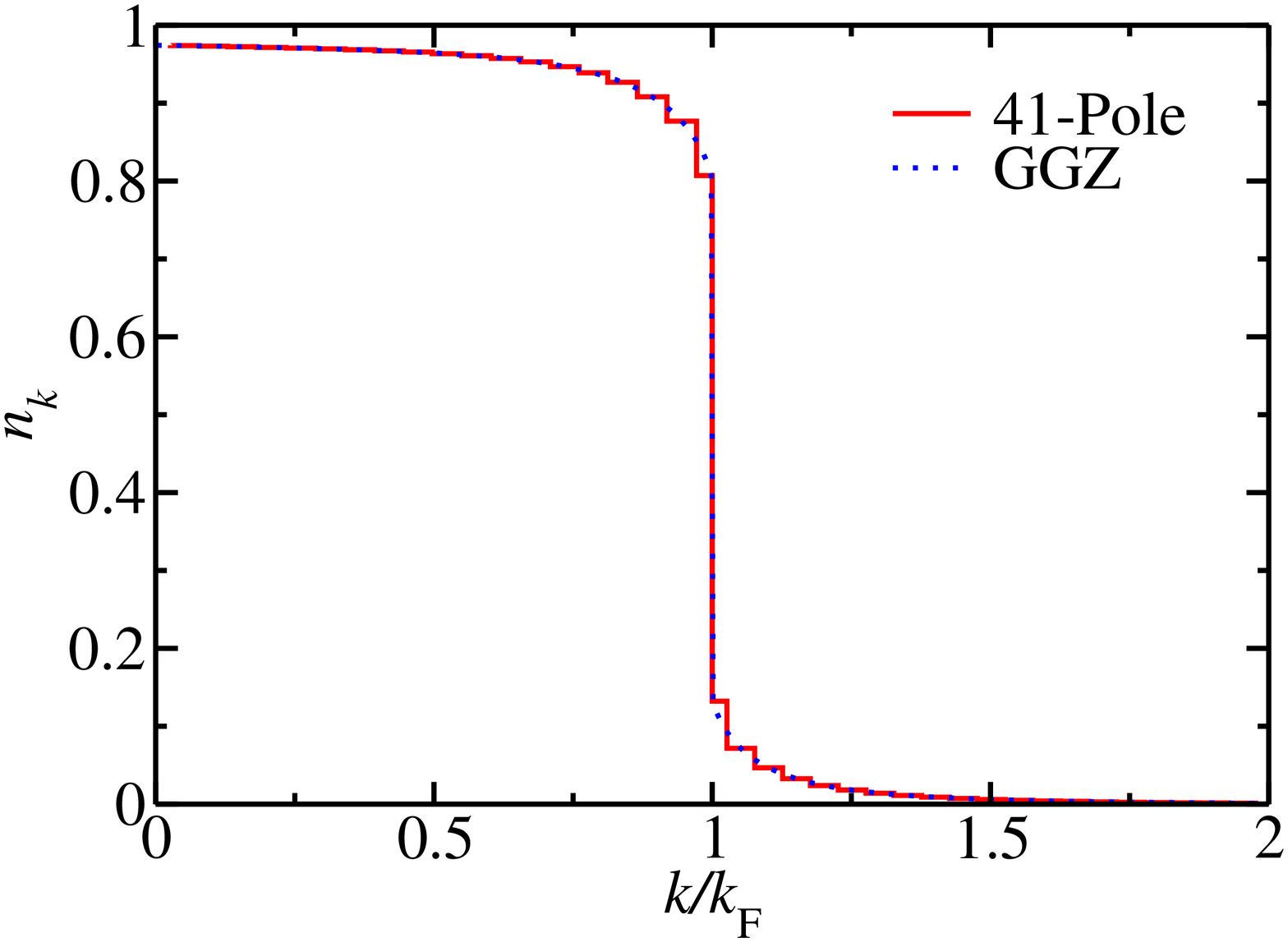}
\caption{\label{fig_41pole_nk}
  Comparison of the momentum distribution function obtained
  within the 41-pole model to the one given in
  Ref.~\cite{PhysRevB.66.235116} (GGZ).
  The dimensionless density-parameter is set to $r_s=3$.
}
\end{figure}

In Fig.~\ref{fig_41pole_nk} we illustrate the basic idea of the
$(n+1)$-pole model by comparing the momentum distribution function
$n_k$ that it produces to the one of Ref.~\cite{PhysRevB.66.235116}:
The bands in the $(n+1)$-pole model of the zero-bandnarrowing approximation
of the UEG are given by the set of parabolas, Eq.~\eqref{eq_ueg_parabolas}.
$n$ of these bands cut the Fermi level. When one of these bands cuts the
Fermi level at a given $k$, i.e., when $E_{k,i}=E_{\rm F}$,
the momentum distribution is reduced by the weight $a_i$
of this band. As Fig.~\ref{fig_41pole_nk}
shows, with $n=40$ the model of Ref.~\cite{PhysRevB.66.235116} can be
reproduced reasonably well. One may of course increase the number of poles
further, until these two curves become indistinguishable.
However, the effect of increasing the number of poles on $M^{(2+)}_{k_{\rm F}}$ and
the charge response $P_{0}$ is small, as Fig.~\ref{fourpole_compare_secmom}
and Fig.~\ref{fig_charge_polari}
show.
Thus, reproducing $Z$, $N_{<}$, and $N_{>}$ by the 4-pole approximation
is already a major improvement over KS-DFT, while increasing the number of
poles further mainly improves $n_{k}$ below and above the Fermi surface,
as the comparison between Fig.~\ref{fig_4pole_nk}
and Fig.~\ref{fig_41pole_nk}
shows.

Interestingly, we may even take
the limit $n\rightarrow \infty$ and consider a continuum model.
In the continuum model the energy of the highest band is
\bege
\begin{aligned}
&E_{\infty}=\frac{1}{n_{\infty}}
\Biggl[
  \int_{0}^{k_{\rm F-}}
  \left(
  E_{\rm KS}+\frac{\hbar^2}{2m}(k_{\rm F}^2-k^2)
  \right)
  \frac{dn_{k}}{dk}dk\\
&+  \int_{k_{\rm F+}}^{\infty}
  \left(
  E_{\rm KS}+\frac{\hbar^2}{2m}
  (k_{\rm F}^2-k^2)
  \right)
  \frac{dn_{k}}{dk}dk-E_{\rm KS}Z
  +M^{(1)}_{k_{\rm F}}\Biggr],
\end{aligned}
\ee
which may be rewritten as
\bege
\begin{aligned}
&E_{\infty}=\frac{1}{n_{\infty}}
  \Biggl[
    -E_{\rm KS}Z
    +M^{(1)}_{k_{\rm F}}-n_{0}\left(
    E_{\rm KS}+\frac{\hbar^2 k_{\rm F}^2}{2m}
    \right)
    \\
  &-\int_{0}^{k_{\rm F-}}
  \frac{\hbar^2}{2m}k^2
  \frac{dn_{k}}{dk}dk
  -  \int_{k_{\rm F+}}^{\infty}
  \frac{\hbar^2}{2m}
  k^2
  \frac{dn_{k}}{dk}dk\Biggr],
\end{aligned}
\ee
where
\bege
n_{\infty}=1-n_{0}.
\ee

The $I$-th moment is given by
\bege
\begin{aligned}
M^{(I)}&=n_{\infty}E_{\infty}^{I}+ZE_{\rm KS}^{I}-
\int_{0}^{k_{\rm F-}}\frac{dn_{k}}{dk}\mathcal{E}^{I}_{k}dk\\
&-\int_{k_{\rm F+}}^{\infty}\frac{dn_{k}}{dk}\mathcal{E}^{I}_{k}dk,
\end{aligned}
\ee
where
\bege
\mathcal{E}_{k}=E_{\rm KS}+\frac{\hbar^2}{2m}
\left[
  k_{\rm F}^2-k^2
  \right].
\ee

\subsection{An optimized two-pole model}
\label{sec_opti_2p}
In weakly and moderately correlated systems the low-energy
satellite band is often of little interest, while a good description
of bandgaps and bandwidths is desirable.
In this case,
the most important benefits of the four-pole approximation
may be reproduced effectively by an optimized two-pole model.
In this model we leave away the low-energy satellite band
completely.
We assume that at the Fermi surface we have $E_1=E_{\rm KS}$ (Seitz-theorem).
For $a_1$ we may either choose $n_{k_{\rm F-}}$ or $Z$. It might be that one of these
two possible options is better, which we have not investigated systematically yet.
The argument to use $a_1=Z$ is that the weight of the quasiparticles on the
Fermi surface at $k_F$ is $Z$ in the 4-pole approximation.
However, one may also argue that $a_1=n_{k_{\rm F-}}$ might be more correct, because the
weight of the low-energy satellite band should be included into $a_1$, if it is
not described explicitly. In this work we choose $a_1=Z$.

We assume that the band $E_1$ cuts the Fermi level at $k_{\rm T}>k_{\rm F}$,
because this is the only way to achieve normalization of $n_{k}$ when the
low-energy satellite band is not described explicitly.
In general both bands, $E_{1}$ and $E_{2}$, may cut the Fermi energy.
When both bands cut the Fermi level
\bege\label{eq_kt_both}
a_2\left[
k_{\rm F}^3-a_{1} k_{\rm T}^3
\right]^2
=\left[
  k_{\rm F}^2
  +V^{\rm c}
  -a_1 k_{\rm T}^2
\right]^3  
\ee
is a nonlinear equation for $k_{\rm T}$, which can be solved
numerically.
Here, $a_2=1-a_1$, and
\bege
V^{c}=\frac{d }{d n_{\rm e}}
\left[
  n_{\rm e}
  E^{\rm c}(n_{\rm e})
  \right]
\ee
is the correlation potential,
where $E^{\rm c}(n_{\rm e})$ is the
correlation energy per particle in the UEG with electron density $n_{\rm e}$.

In practice, one may first solve Eq.~\eqref{eq_kt_both}
at all relevant parameters $r_s$.
Next, at a given $r_s$ one checks
if
\bege
E_2-\frac{\hbar^2 k_{\rm T}^2}{2m}>E_{\rm KS}
\ee
is satisfied. If it is, the band $E_2$ does not cut the Fermi energy.
In these cases, one needs to replace
$k_{\rm T}$ by
\bege
k_{\rm T}=\frac{k_{\rm F}}{a_1^{1/3}}.
\ee

The energy of the second band is computed from
\bege
E_2=\frac{\bar{M}^{(1)}_{k_{\rm F}}-a_1 E_{\rm KS}}{a_2},
\ee
where
\bege
\bar{M}^{(1)}_{k_{\rm F}}=\frac{\hbar^2 k_{\rm T}^2}{2m}
-
\frac{\hbar^2}{m a_{\rm B}^2}
\left[
\frac{3}{2\pi}
  \right]^{\frac{2}{3}}
\frac{1}{r_s}.
\ee

Ignoring the low-energy satellite bands according to these
equations affects the determination of the Fermi energy:
In insulators one will automatically include part of the
conduction band electrons
into the ground state charge density
in order to achieve charge neutrality.
In the calculation of the DOS, this problem can be avoided
simply by setting $a_{j}=1$ in the determination of the Fermi level.
By this choice one assumes that the missing charge is
provided by the satellite bands and amounts effectively to
\bege
a_{j}\rightarrow 1.
\ee
When computing the charge density in the self-consistency
loop we could in principle use $a_{j}$ as obtained from the
state vectors~\cite{momentis}.
However, in order to minimize the inconsistency between these
two options, we use
\bege
a_{j}\rightarrow
\frac{1}{2}
\left[
1+a_{j}
  \right]
\ee
in the selfconsistency loop.
\section{Results}
\label{sec_results}
\begin{figure}
\includegraphics[angle=0,width=\linewidth]{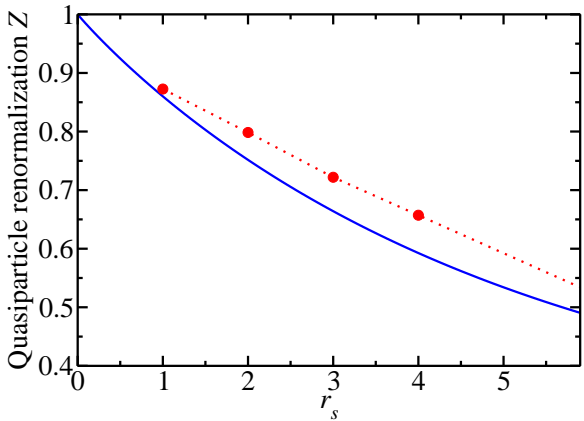}
\caption{\label{fig_zfactor_compare}
  Quasiparticle renormalization factor $Z$ as obtained from the
  model of Ref.~\cite{PhysRevB.66.235116} (solid line) and
  from the variational diagrammatic Monte Carlo calculations of Ref.~\cite{Haule2022_diagrammatic} (circles).
  Dotted lines
  are linear interpolations and extrapolations of the variational diagrammatic
  Monte Carlo data.
}
\end{figure}

Unless stated otherwise, the results shown in this section are obtained
with the optimized two-pole model of Sec.~\ref{sec_opti_2p}.
An important ingredient of this model is the quasiparticle renormalization
factor $Z$, for which many calculations have been
performed~\cite{PhysRevLett.107.110402,Haule2022_diagrammatic}.
Ref.~\cite{Haule2022_diagrammatic} reports a very small error
bar for recent variational diagrammatic Monte Carlo
computations of this quantity. In Fig.~\ref{fig_zfactor_compare}
we compare these recent results to the
model of Ref.~\cite{PhysRevB.66.235116}.
At the parameters $r_s=2$, $r_s=3$, and $r_s=4$ the deviations are large,
while the agreement is good at $r_s=1$.

We need a model for $Z(r_s)$ which
captures the full range of variation of $r_s$ in our first-principles calculations.
The minimal and maximal values of $r_s$ are listed in Table~\ref{tab_rsminmax}
for all systems studied in this work. Clearly, we need $Z$ also for values
below and above the range considered in Ref.~\cite{Haule2022_diagrammatic},
which lists values of $Z$
only for $r_s=1$, $r_s=2$, $r_s=3$, and $r_s=4$. Therefore, we construct a model
of $Z(r_s)$ as follows: For $r_s<1$ we use the model of Ref.~\cite{PhysRevB.66.235116}. This is
justified, because at $r_s=1$ this model does not deviate much from the variational diagrammatic
Monte Carlo results of Ref.~\cite{Haule2022_diagrammatic} (see Fig.~\ref{fig_zfactor_compare}).
The variational diagrammatic
Monte Carlo results in the range $1\le r_s\le4$ almost follow a linear
trend. Therefore, we use linear interpolation to determine  $Z(r_s)$ for values
in the range $1< r_s<4$ (dotted lines in Fig.~\ref{fig_zfactor_compare}).
According to Table~\ref{tab_rsminmax} we need $Z(r_s)$ up to $r_s=5.07$.
Therefore, we linearly extrapolate the variational diagrammatic Monte Carlo results for $r_s>4$.

\subsection{Silicon, Diamond, and Silicon Carbide}
\begin{figure}
\includegraphics[angle=0,width=\linewidth]{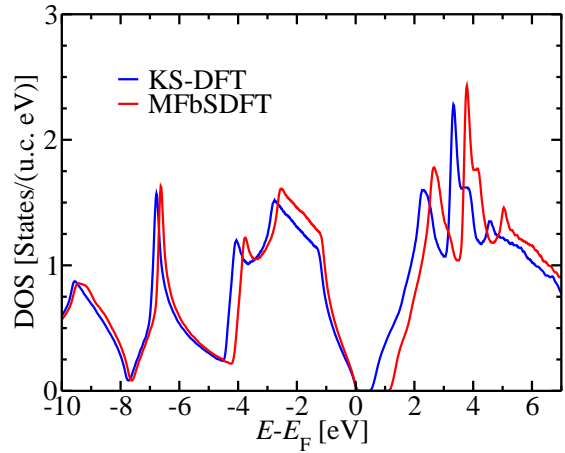}
\caption{\label{DOS_Si}   Density of states (DOS)
  of Si vs.\ energy $E$
  as obtained in KS-DFT and in MFbSDFT. $E_{\rm F}$ is the
  Fermi energy.
}
\end{figure}

Silicon crystallizes in the diamond structure with the
lattice parameter $a=5.43$~\AA.
Employing the PBE~\cite{PhysRevLett.77.3865}
functional we obtain a bandgap
of 0.6~eV within KS-DFT. This is smaller than the
experimental bandgap of 1.17~eV by roughly a factor of 2.
In Fig.~\ref{DOS_Si} we compare the DOS
obtained from MFbSDFT 
to the one obtained from KS-DFT. Within MFbSDFT the
bandgap is 1.22~eV, which is close to the experimental value.

\begin{figure}
\includegraphics[angle=0,width=\linewidth]{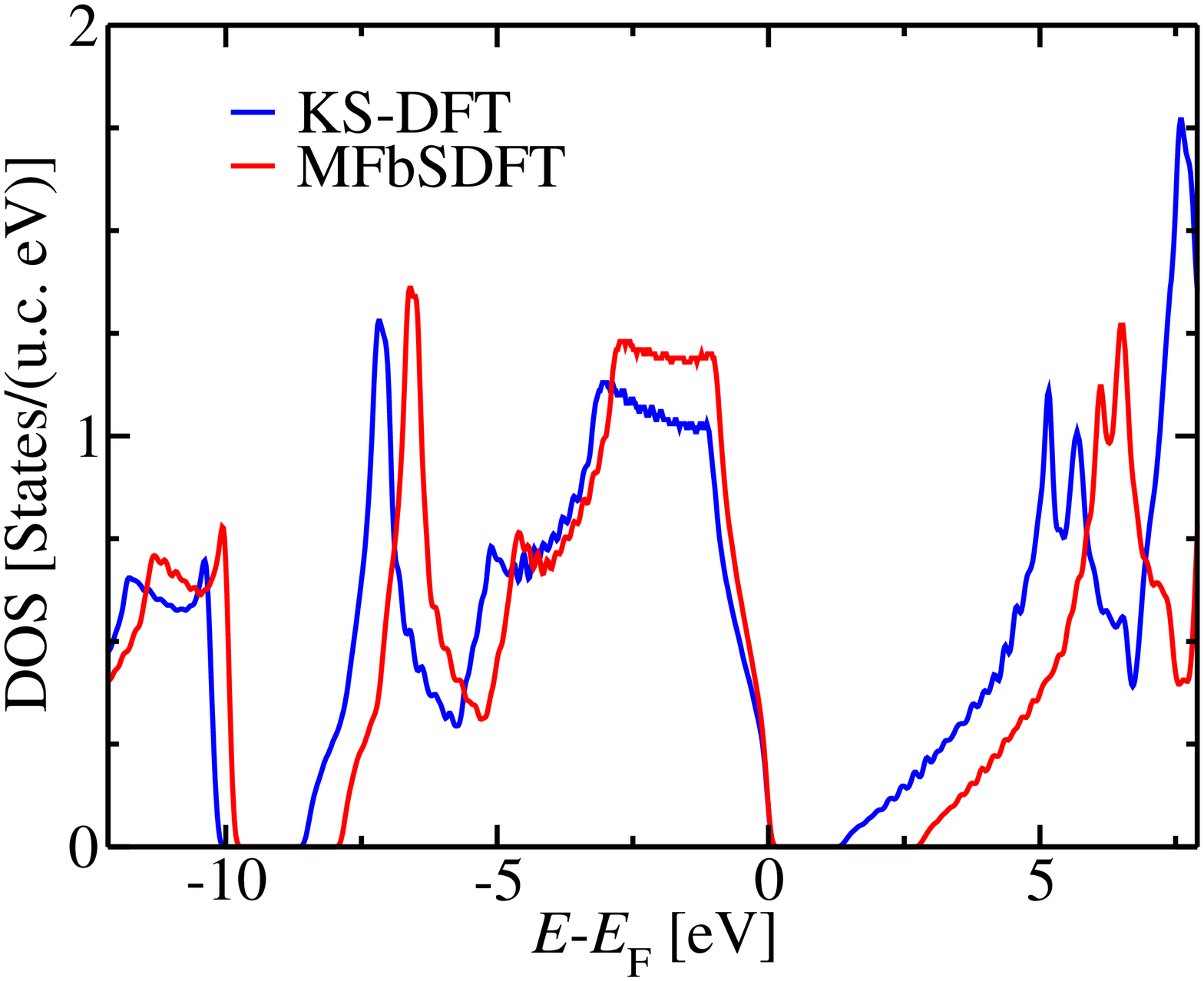}
\caption{\label{DOS_SiC} Density of states (DOS)
  of SiC vs.\ energy $E$
  as obtained in KS-DFT and in MFbSDFT. $E_{\rm F}$ is the
  Fermi energy.
}
\end{figure}

In the cubic 3C-SiC polymorph of silicon carbide
one half of the sites of the
diamond lattice are occupied by Si and the other half by C.
The lattice constant is $a=4.36$~\AA.
Experimentally, the band gap is determined to be 2.36~eV.
Employing the PBE functional we obtain a bandgap of 1.4~eV in KS-DFT.
The bandgap of 2.8~eV obtained within MFbSDFT is too large compared
to the experiment, but significantly closer to the experimental value
than the PBE result.
In Fig.~\ref{DOS_SiC} we compare the DOS
obtained from MFbSDFT 
to the one obtained from KS-DFT.

\begin{figure}
\includegraphics[angle=0,width=\linewidth]{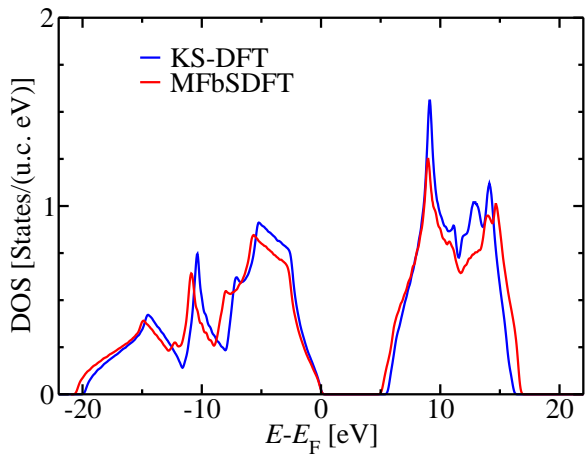}
\caption{\label{DOS_C} Density of states (DOS)
  of diamond vs.\ energy $E$
  as obtained in KS-DFT and in MFbSDFT. $E_{\rm F}$ is the
  Fermi energy.
}
\end{figure}

The lattice parameter of diamond is $a=3.567$~\AA.
Employing the PBE functional we obtain a bandgap of 5.7~eV in KS-DFT,
which is in good agreement with the experimental bandgap of 5.47~eV.
Within MFbSDFT we obtain a bandgap of 5.14~eV, which also agrees acceptably
well with experiment, while the agreement is slightly better for the PBE result.
In Fig.~\ref{DOS_C} we compare the DOS
obtained from MFbSDFT 
to the one obtained from KS-DFT.

It is remarkable that KS-DFT almost reproduces the experimental bandgap in diamond,
but is significantly off in the isoelectric compounds SiC and Si. To some extent this
depends on the functional used. For example KS-DFT predicts a 
bandgap of only 4.11~eV when LDA is employed.
However, this underestimation of the bandgap by 25\% is still relatively
small compared to the typical bandgap error in KS-DFT.

Since an important contribution to
the moment functional corrections comes from the quasiparticle renormalization factor $Z$,
the question arises of whether this correction might be particularly small in diamond,
which might contribute to the good performance of PBE for the bandgap in this wide-gap
semiconductor.
According
to Fig.~\ref{fig_zfactor_compare} the $Z$ factor starts to deviate strongly from 1
when $r_s$ becomes larger than 1. Therefore, we show in Table~\ref{tab_rsminmax}
the minimal and maximal values of $r_s$. Indeed, diamond is characterized by a relatively small
value of $r_{s,{\rm max}}=2.44$. Thus, to some extent Table~\ref{tab_rsminmax} suggests 
that the importance of taking $Z<1$ into account when $r_s$ increases is reflected in the
error $\Delta E^{\rm gap}_{\rm PBE}/E^{\rm gap}_{\rm exp}$.
A more quantitative investigation of this hypothesis might be possible
by computing the unit-cell average
\bege
\frac{1}{V}
\int
d^{3}r
\left[
1-Z(r_s)
\right],
\ee
which we do not consider here and leave for future work.

\begin{threeparttable}
\caption{
  Minimal ($r_{s,{\rm min}}$) and maximal ($r_{s,{\rm max}}$) values of the
  dimensionless density parameter $r_s$
  in the MFbSDFT calculations.
  Deviation $\Delta E^{\rm gap}_{\rm PBE}=E^{\rm gap}_{\rm PBE}-E^{\rm gap}_{\rm exp}$
  of the bandgap obtained with KS-DFT ($E^{\rm gap}_{\rm PBE}$) from the experimental
  bandgap ($E^{\rm gap}_{\rm exp}$).
  Deviation $\Delta E^{\rm gap}_{\rm SDFT}=E^{\rm gap}_{\rm SDFT}-E^{\rm gap}_{\rm exp}$
  of the bandgap obtained with MFbSDFT ($E^{\rm gap}_{\rm SDFT}$) from the experimental
  bandgap ($E^{\rm gap}_{\rm exp}$). Large relative
  errors $|\Delta E^{\rm gap}_{\rm PBE}/E^{\rm gap}_{\rm exp}|$
  tend to occur when $r_{s,{\rm max}}$ is large.
}
\label{tab_rsminmax}
\begin{ruledtabular}
\begin{tabular}{c|c|c|c|c|}
&$r_{s,{\rm min}}$
&$r_{s,{\rm max}}$
  &$\Delta E^{\rm gap}_{\rm PBE}/E^{\rm gap}_{\rm exp}$
 &$\Delta E^{\rm gap}_{\rm SDFT}/E^{\rm gap}_{\rm exp}$ 
  \\
  \hline
C
&0.12
&2.44
&4\%
&-6\%\\
\hline
Si
&0.05
&4.28
&-49\%
&4\%\\
\hline
SiC
&0.05
&3.29
&-41\%
&19\%\\
\hline
BN
&0.1
&5.07
&-29\%
&-6\%\\
\hline
MgO
&0.059
&2.72
&-38\%
&-11\%\\
\hline
CaO
&0.03
&3.32
&-48\%
&-13\%\\
\hline
ZnO
&0.019
&3.72
&-73\%
&2\%\\
\end{tabular}
\end{ruledtabular}
\end{threeparttable}

\subsection{Boron nitride}
\begin{figure}
\includegraphics[angle=-90,width=\linewidth]{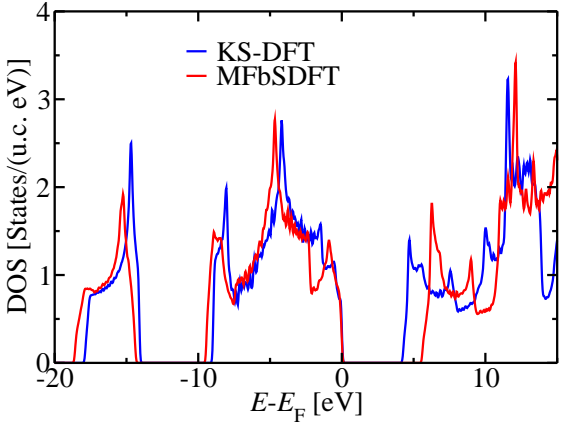}
\caption{\label{BN_DOS} Density of states (DOS)
   of BN vs.\ energy $E$
  as obtained in KS-DFT and in MFbSDFT. $E_{\rm F}$ is the
  Fermi energy.
}
\end{figure}

We consider hexagonal BN, which exhibits a layered
structure similar to graphite. It is a 
semiconductor with a wide bandgap of 5.955~eV~\cite{Cassabois2016}.
The in-plane lattice parameter is $a=2.504$~\AA\,
and the interlayer distance is $a=3.33$~\AA.
KS-DFT with the PBE functional significantly underestimates
the bandgap and predicts it to be 4.25~eV.
In contrast, the bandgap obtained from MFbSDFT is 5.61~eV, which is
in good agreement with experiment.
In Fig.~\ref{BN_DOS}
we compare the DOS of the MFbSDFT calculation
to the one obtained within KS-DFT.

\subsection{MgO and CaO}
\begin{figure}
\includegraphics[angle=0,width=\linewidth]{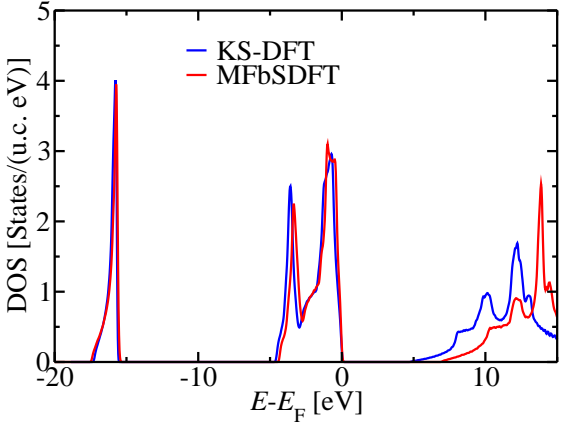}
\caption{\label{DOS_MgO} Density of states (DOS)
   of MgO vs.\ energy $E$
  as obtained in KS-DFT and in MFbSDFT. $E_{\rm F}$ is the
  Fermi energy.
}
\end{figure}

MgO crystallizes in the rocksalt structure with the
lattice parameter $a=4.212$~\AA.
Using KS-DFT with the PBE functional we obtain a
bandgap of 4.8~eV, which is much smaller than the
experimental bandgap of 7.77~eV.
In MFbSDFT we obtain a bandgap of 6.95~eV.
This is in acceptable agreement with experiment considering
that GW calculations deviate from the experimental bandgap
as well in this case even when quasiparticle self-consistency is
imposed~\cite{PhysRevLett.96.226402}. 
In Fig.~\ref{DOS_MgO}
we compare the DOS of the MFbSDFT calculation
to the one obtained within KS-DFT.

\begin{figure}
\includegraphics[angle=0,width=\linewidth]{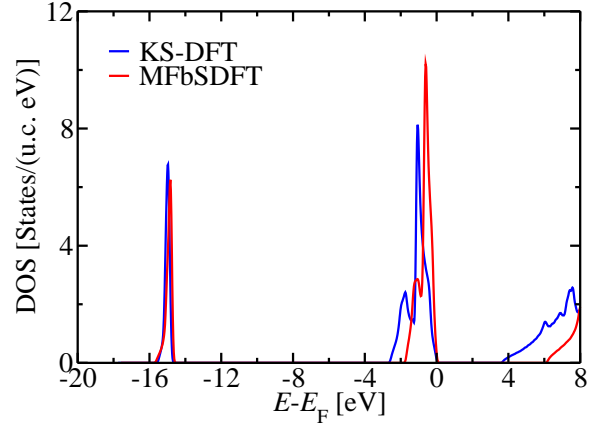}
\caption{\label{DOS_CaO} Density of states (DOS)
  of CaO vs.\ energy $E$
  as obtained in KS-DFT and in MFbSDFT. $E_{\rm F}$ is the
  Fermi energy.
}
\end{figure}

The isoelectric compound CaO
crystallizes in the rocksalt structure with the
lattice parameter $a=4.815$~\AA.
Using KS-DFT with the PBE functional we obtain a
bandgap of 3.67~eV, which is much smaller than the
experimental bandgap of 7.1~eV.
In MFbSDFT we obtain a bandgap of 6.17~eV, which is a significant
improvement.
In Fig.~\ref{DOS_CaO}
we compare the DOS of the MFbSDFT calculation
to the one obtained within KS-DFT.

\subsection{ZnO}
\begin{figure}
\includegraphics[angle=0,width=\linewidth]{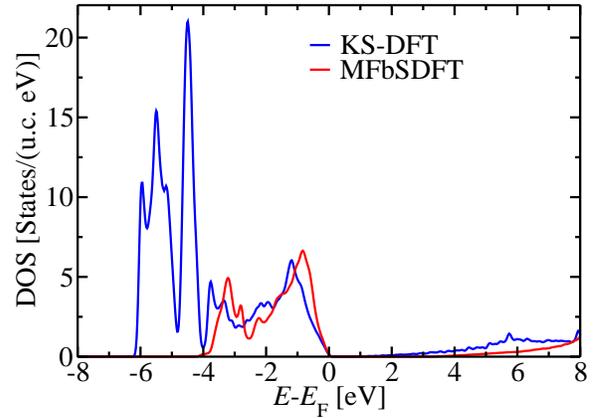}
\caption{\label{DOS_ZnO} Density of states (DOS)
  of ZnO vs.\ energy $E$
  as obtained in KS-DFT and in MFbSDFT. $E_{\rm F}$ is the
  Fermi energy.
}
\end{figure}

ZnO crystallizes in the wurtzite crystal structure.
The lattice parameters are $a=3.25$~\AA\,
and $c=5.2$~\AA.
When we employ the PBE functional we obtain a
bandgap of 0.88~eV from KS-DFT, which is much smaller than the
experimental bandgap of 3.3~eV.
In MFbSDFT we obtain a bandgap of 3.37~eV, which is very close to the
experimental value.
In Fig.~\ref{DOS_ZnO}
we compare the DOS of the MFbSDFT and of the
KS-DFT calculations.

\subsection{The optimized two-pole model vs.\ the three-pole approximation}
In Ni correlation effects are very important and standard KS-DFT
fails to predict the correct exchange splitting, the correct bandwidth,
and the valence-band satellite peak~\cite{nickel_PhysRevB.40.5015,nickel_Borgiel1990}.
Our optimized two-pole model of Sec.~\ref{sec_opti_2p}
excludes the low-energy satellite band by construction.
At the same time it is possible to obtain the satellite
peak in Ni using the first four spectral
moments obtained
from a lattice model~\cite{nickel_PhysRevB.40.5015,nickel_Borgiel1990}.
The question therefore arises of whether our prescription
in Sec.~\ref{sec_opti_2p}
to
construct the moment potentials for MFbSDFT suppresses the
valence band satellite in Ni.

\begin{figure}
\includegraphics[angle=-90,width=\linewidth]{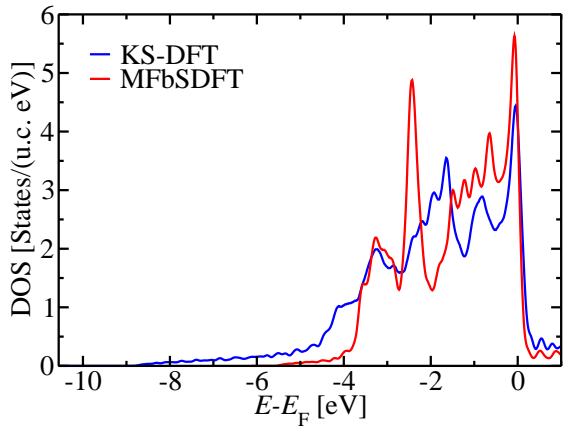}
\caption{\label{fig_ni_dos_2pole}
  Density of states (DOS) in the spin-unpolarized phase of Ni.
  KS-DFT and MFbSDFT (optimized 2-pole model) are compared.
}
\end{figure}

In Fig.~\ref{fig_ni_dos_2pole}
we show that the valence band satellite is indeed absent
when the moment potentials are constructed from the model of Sec.~\ref{sec_opti_2p}.
However, the comparison to KS-DFT shows that at least the
bandwidth is smaller in MFbSDFT and therefore in better agreement with experiment.
Since we do not consider the spin-polarized case in this work,
Fig.~\ref{fig_ni_dos_2pole} illustrates the DOS of a spin-unpolarized phase of Ni.
This interesting result shows that the presence or absence of a valence band
satellite peak in the spectrum is not only dependent on the number of poles used,
but also on the prescription used to obtain the moment potentials.

\begin{figure}
\includegraphics[angle=0,width=\linewidth]{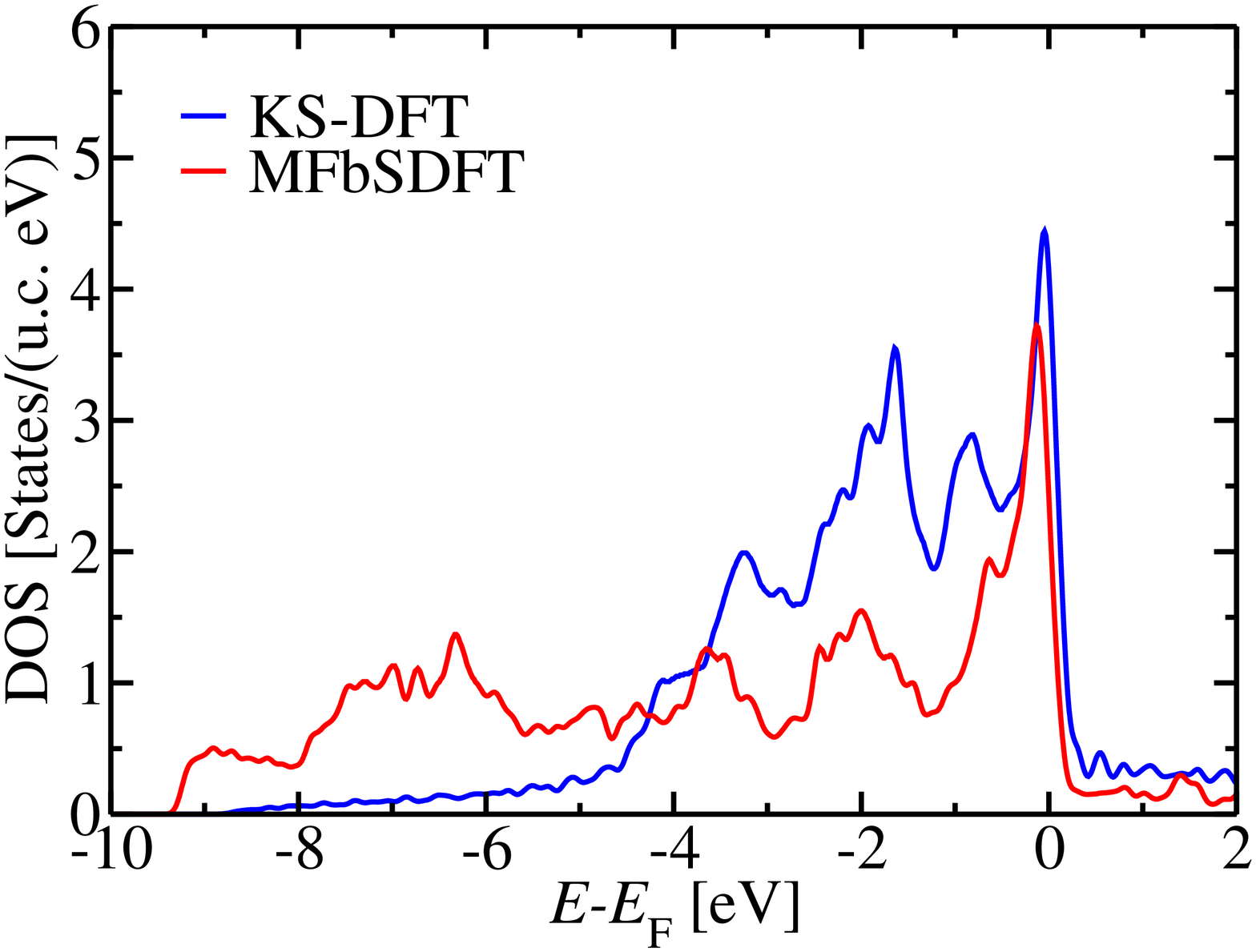}
\caption{\label{fig_ni_dos_3pole}
  Density of states (DOS) in the spin-unpolarized phase of Ni.
  KS-DFT and MFbSDFT (3-pole model) are compared.
}
\end{figure}

In Sec.~\ref{sec_threepole} we have discussed that the
3-pole model gives too much weight to the satellite band.
This can be seen clearly in Fig.~\ref{fig_ni_dos_3pole}.
While the experimentally observed satellite peak slightly below -6~eV is
now present in the MFbSDFT spectrum, there is overall too much spectral
density below -4~eV relative to the main band.
A possible solution might be to optimize the 3-pole model using part of the
ideas of Sec.~\ref{sec_opti_2p}, which we leave for future work.

  On the one hand the valence-band satellite in spin-unpolarized Ni has been
  investigated before theoretically, on the other hand the
  ultimate test of a new theoretical approach is the comparison
  to experiment. The question of the existence of valence band
  satellites in the spectra of spin-polarized Ni, Fe, and Co
  has been given a lot of attention theoretically and
  experimentally~\cite{nickel_PhysRevB.40.5015,bcc_iron_Nolting1995,PhysRevB.36.887,PhysRevB.55.6678,PhysRevB.61.12582,PhysRevB.70.233103,PhysRevB.85.205109},
  while little data are available for the spin-unpolarized phases of
  these materials. Unfortunately, there is also very little data in the literature
  on the momentum-distribution function in the spin-polarized UEG.
  An important task left for future work is therefore the development
  of reliable models of the $Z$ renormalization and of the momentum
  distribution function in the spin-polarized UEG. Once these are
  available, one may develop a spin-polarized MFbSDFT and compare
  the MFbSDFT spectra of spin-polarized Ni, Fe, and Co
  with the experimental ones.

\section{Summary}
\label{sec_summary}
Considering the 3-pole and the 4-pole approximations of the
spectral function of the UEG we corroborate the idea
that more and more
properties of the UEG can be described when the number of poles is
increased. Our 4-pole model describes the charge response, the
momentum distribution, and the second moment of the UEG acceptably well,
while the KS-DFT approximation of the UEG describes from these quantities
only the charge response
well. Focusing on the most important aspects that the 4-pole approximation
improves in comparison to KS-DFT in weakly and moderately correlated systems
we construct an optimized 2-pole approximation
which we use to extract \textit{parameter-free universal} moment potentials for MFbSDFT.
Using these we show that the bandgaps in the insulators 
Si, SiC, BN, MgO, CaO, and ZnO are significantly closer to the experimental
bandgaps in MFbSDFT than in KS-DFT with the PBE functional.
Finally, we show that the 3-pole approximation highly overestimates the satellite
bands in strongly correlated materials and therefore it needs to be optimized 
to mimic the behaviour of the 4-pole approximation.
In this work we consider only the spin unpolarized case. The inclusion of magnetism
is an important extension left for future work.

\appendix
\section{Model of the second moment for the UEG}
\label{app_model_secmom}
Ref.~\cite{PhysRevB.69.045113}
derives a model for the second spectral moment of the UEG,
where
$M^{(2)}_{k}$ is expressed in terms of the
exchange self-energy, the
pair correlation function, and a remaining higher correlation function,
which is assessed by the single-Slater-determinant approximation.
From this model we obtain
\bege\label{eq_m2p_heg}
M^{(2+)}_{k}=\Sigma^{(1)}_{\rm loc}+\Sigma^{(1)}_{{\rm nl},k},
\ee
where
\bege\label{eq_sigmaloc_heg}
\Sigma^{(1)}_{\rm loc}=
\left[
\frac{\hbar^2}{2 m a^2_{\rm B}}
\right]^2
\frac{32}{3\pi^2}\frac{k_{\rm F}}{(\bar{\alpha} r_s)^2}
\int_{0}^{\infty}\frac{d\,k'}{k'^2}S(k')
\ee
is a $k$-independent contribution~\cite{PhysRevB.69.045113},
and 
\bege\label{eq_sigmanlk_heg}
\begin{aligned}
  &\Sigma^{(1)}_{{\rm nl},k}=
-\left[
\Sigma_{k}^{(0)}
\right]^2
-
\left[
\frac{\hbar^2}{2 m a_{\rm B}^2}
\right]^2
\frac{2}{(\bar{\alpha} r_{s})^3 \pi k k_{\rm F}}\times\\
&\times\int_{0}^{\infty}d\,k'\,k'\,\Sigma_{k'}^{(0)}(1-2n_{k'})
\ln
\left|
\frac{k-k'}{k+k'}
\right|
\end{aligned}
\ee
is a $k$-dependent contribution~\cite{PhysRevB.69.045113}.
Here, $\bar{\alpha}=[4/(9\pi)]^{1/3}$, $k_{\rm F}=(\bar{\alpha} r_s a_{\rm B})^{-1}$,
\bege\label{eq_struc_fac_heg}
S(k)=1+4\pi n_{\rm e}\int d\,r r
\frac{\sin(kr)}{k}[g(r)-1],
\ee
is the structure factor,
$g(r)$ is the pair correlation function,
and
\bege\label{eq_sigma0_heg}
\Sigma_{k}^{(0)}=-
\frac{\hbar^2}{2 m a^2_{\rm B}}
\frac{2}{\pi k k_{\rm F}\bar{\alpha} r_s}
\int_{0}^{\infty}
d\,k'\,k' \,n_{k'}
\ln
\left|
\frac{k+k'}{k-k'}
\right|
\ee
is the exchange self-energy.

In order to extract the moment potential $\mathcal{V}^{(2+)}$ from this model
we set $k=k_{\rm F}$, because we need $k$-independent moment potentials. Thus,
we use
\bege
\mathcal{V}^{(2+)}=\Sigma^{(1)}_{\rm loc}+\Sigma^{(1)}_{{\rm nl},k_{\rm F}}.
\ee
$\mathcal{V}^{(2+)}$ depends on $r_s$ because
Eq.~\eqref{eq_sigmaloc_heg}
and Eq.~\eqref{eq_sigmanlk_heg}
depend explicitly on $r_s$.
Additionally, it depends on $r_s$ because $k_{\rm F}$ and $S(k)$
depend on $r_s$.
In the literature several models have been suggested for the pair correlation function $g(r)$
and for the structure factor $S(k)$. We use the model of Ref.~\cite{PhysRevB.61.7353}
in the numerical results shown in Fig.~\ref{fourpole_compare_secmom}.

\section{MFbSDFT with the first 6 spectral moments}
\label{sec_app_mfbsdft_6mom}
In Ref.~\cite{momdis}
we describe an efficient algorithm to obtain the spectral
function from the first $2P$ spectral moment matrices of size $N\times N$.
In the special case of $P=3$, i.e., when the first 6
spectral moment matrices
$\vn{M}^{(0)}_{\vn{k}}$,
$\vn{M}^{(1)}_{\vn{k}}$,
$\vn{M}^{(2)}_{\vn{k}}$,
$\vn{M}^{(3)}_{\vn{k}}$,
$\vn{M}^{(4)}_{\vn{k}}$,
$\vn{M}^{(5)}_{\vn{k}}$
are used,
the $3N$
poles $E_{\vn{k},j}$ ($j=1,\dots,3N$) of the spectral function
are given by the eigenvalues of the $3N\times 3N$ matrix
\bege
\vn{\mathcal{H}}_{\vn{k}}=
\begin{pmatrix}
  \vn{M}^{(1)}_{\vn{k}} &\vn{B}_{1,\vn{k}}\\
  \vn{B}_{1,\vn{k}}^{\dagger} &\vn{D}_{1,\vn{k}}
\end{pmatrix},
\ee
where
$\vn{B}_{1,\vn{k}}$ is a $N\times 2N$ matrix,
and $\vn{D}_{1,\vn{k}}$ is a $2N\times 2N$ matrix.
$\vn{B}_{1,\vn{k}}$ is given by the first $N$ rows of the
$2N\times 2N$ matrix $\mathscrbf{B}$, which may be
computed by taking the square root of the
hermitian positive definite matrix
\bege\label{eq_bbdag_general}
\mathscrbf{B}\mathscrbf{B}^{\dagger}=
\begin{pmatrix}
  \vn{M}_{\vn{k}}^{(2)}-\left[\vn{M}_{\vn{k}}^{(1)}\right]^2\quad\quad &\vn{M}_{\vn{k}}^{(3)}-\vn{M}_{\vn{k}}^{(1)}\vn{M}_{\vn{k}}^{(2)}\\
  \vn{M}_{\vn{k}}^{(3)}-\vn{M}_{\vn{k}}^{(2)}\vn{M}_{\vn{k}}^{(1)}\quad\quad &\vn{M}_{\vn{k}}^{(4)}-\left[\vn{M}_{\vn{k}}^{(2)}\right]^2\\
\end{pmatrix}.
\ee

The matrix $\vn{D}_{1,\vn{k}}$ may be computed from
\bege\label{eq_d1_general}
\vn{D}_{1,\vn{k}}=
\mathscrbf{B}^{-1}
\begin{pmatrix}
  \vn{B}_{2,\vn{k}}-\vn{M}_{\vn{k}}^{(1)}\vn{B}_{1,\vn{k}}\\
  \vn{B}_{3,\vn{k}}-\vn{M}_{\vn{k}}^{(2)}\vn{B}_{1,\vn{k}}
\end{pmatrix},
\ee
where the $N\times 2N$ matrix $\vn{B}_{2,\vn{k}}$
is given by the last $N$
rows of the matrix $\mathscrbf{B}$
and the $N\times 2N$ matrix $\vn{B}_{3,\vn{k}}$ may be computed from
the $2N \times N$ matrix
\bege
\vn{B}_{3,\vn{k}}^{\dagger}=\mathscrbf{B}^{-1}
\begin{pmatrix}
  \vn{M}_{\vn{k}}^{(4)}-\vn{M}_{\vn{k}}^{(1)}\vn{M}_{\vn{k}}^{(3)}\\
  \vn{M}_{\vn{k}}^{(5)}-\vn{M}_{\vn{k}}^{(2)}\vn{M}_{\vn{k}}^{(3)}
\end{pmatrix}
\ee
by taking the complex conjugate.

The single-particle spectral function is given by
\bege\label{eq_final_spec_general}
\frac{S_{\vn{k}ij}(E)}{\hbar}=\sum_{l=1}^{3N}a_{\vn{k}l}\mathcal{V}_{\vn{k}il}\mathcal{V}^{*}_{\vn{k}jl}\delta(E-E_{\vn{k}l}),
\ee
where
\bege\label{eq_statevecs_general}
\mathcal{V}_{\vn{k}i j}=\frac{U_{\vn{k}i j}}{\sqrt{a_{\vn{k}j}}},
\ee
with $i=1,...,N$, $j=1,...,3N$,
is the matrix of state vectors, and
\bege
\label{eq_spec_wei}
a_{\vn{k}j}=\sum_{i=1}^{N} U_{\vn{k}i j} [U_{\vn{k}i j}]^{*}
\ee
are the spectral weights.
The $3N\times 3N$ matrix  $\vn{U}_{\vn{k}}$ holds the
$3N$ eigenvectors of $\vn{\mathcal{H}}_{\vn{k}}$ in its $3N$ columns.

\section*{Acknowledgments}
The project is funded by the Deutsche
Forschungsgemeinschaft (DFG, German Research Foundation) $-$ TRR 288 $-$ 422213477 (project B06),
CRC 1238, Control and Dynamics of Quantum Materials: Spin
orbit coupling, correlations, and topology (Project No. C01),
SPP 2137 ``Skyrmionics",  and Sino-German research project
DISTOMAT (DFG project MO \mbox{1731/10-1}).
We also
acknowledge financial support from the European Research
Council (ERC) under the European Union’s Horizon 2020
research and innovation program (Grant No. 856538, project
``3D MAGiC'') and computing resources granted by the Jülich Supercomputing
Centre under
project No.~jiff40.

\bibliography{insulanis}

\begin{thebibliography}{39}%
\makeatletter
\providecommand \@ifxundefined [1]{%
 \@ifx{#1\undefined}
}%
\providecommand \@ifnum [1]{%
 \ifnum #1\expandafter \@firstoftwo
 \else \expandafter \@secondoftwo
 \fi
}%
\providecommand \@ifx [1]{%
 \ifx #1\expandafter \@firstoftwo
 \else \expandafter \@secondoftwo
 \fi
}%
\providecommand \natexlab [1]{#1}%
\providecommand \enquote  [1]{``#1''}%
\providecommand \bibnamefont  [1]{#1}%
\providecommand \bibfnamefont [1]{#1}%
\providecommand \citenamefont [1]{#1}%
\providecommand \href@noop [0]{\@secondoftwo}%
\providecommand \href [0]{\begingroup \@sanitize@url \@href}%
\providecommand \@href[1]{\@@startlink{#1}\@@href}%
\providecommand \@@href[1]{\endgroup#1\@@endlink}%
\providecommand \@sanitize@url [0]{\catcode `\\12\catcode `\$12\catcode
  `\&12\catcode `\#12\catcode `\^12\catcode `\_12\catcode `\%12\relax}%
\providecommand \@@startlink[1]{}%
\providecommand \@@endlink[0]{}%
\providecommand \url  [0]{\begingroup\@sanitize@url \@url }%
\providecommand \@url [1]{\endgroup\@href {#1}{\urlprefix }}%
\providecommand \urlprefix  [0]{URL }%
\providecommand \Eprint [0]{\href }%
\providecommand \doibase [0]{https://doi.org/}%
\providecommand \selectlanguage [0]{\@gobble}%
\providecommand \bibinfo  [0]{\@secondoftwo}%
\providecommand \bibfield  [0]{\@secondoftwo}%
\providecommand \translation [1]{[#1]}%
\providecommand \BibitemOpen [0]{}%
\providecommand \bibitemStop [0]{}%
\providecommand \bibitemNoStop [0]{.\EOS\space}%
\providecommand \EOS [0]{\spacefactor3000\relax}%
\providecommand \BibitemShut  [1]{\csname bibitem#1\endcsname}%
\let\auto@bib@innerbib\@empty
%</preamble>
\bibitem [{\citenamefont {Vosko}\ \emph {et~al.}(1980)\citenamefont {Vosko},
  \citenamefont {Wilk},\ and\ \citenamefont {Nusair}}]{vwn}%
  \BibitemOpen
  \bibfield  {author} {\bibinfo {author} {\bibfnamefont {S.~H.}\ \bibnamefont
  {Vosko}}, \bibinfo {author} {\bibfnamefont {L.}~\bibnamefont {Wilk}},\ and\
  \bibinfo {author} {\bibfnamefont {M.}~\bibnamefont {Nusair}},\ }\bibfield
  {title} {\bibinfo {title} {Accurate spin-dependent electron liquid
  correlation energies for local spin density calculations: a critical
  analysis},\ }\href {https://doi.org/10.1139/p80-159} {\bibfield  {journal}
  {\bibinfo  {journal} {Canadian Journal of Physics}\ }\textbf {\bibinfo
  {volume} {58}},\ \bibinfo {pages} {1200} (\bibinfo {year}
  {1980})}\BibitemShut {NoStop}%
\bibitem [{\citenamefont {Perdew}\ and\ \citenamefont
  {Wang}(1992)}]{PhysRevB.45.13244}%
  \BibitemOpen
  \bibfield  {author} {\bibinfo {author} {\bibfnamefont {J.~P.}\ \bibnamefont
  {Perdew}}\ and\ \bibinfo {author} {\bibfnamefont {Y.}~\bibnamefont {Wang}},\
  }\bibfield  {title} {\bibinfo {title} {Accurate and simple analytic
  representation of the electron-gas correlation energy},\ }\href
  {https://doi.org/10.1103/PhysRevB.45.13244} {\bibfield  {journal} {\bibinfo
  {journal} {Phys. Rev. B}\ }\textbf {\bibinfo {volume} {45}},\ \bibinfo
  {pages} {13244} (\bibinfo {year} {1992})}\BibitemShut {NoStop}%
\bibitem [{\citenamefont {Haule}\ and\ \citenamefont
  {Chen}(2022)}]{Haule2022_diagrammatic}%
  \BibitemOpen
  \bibfield  {author} {\bibinfo {author} {\bibfnamefont {K.}~\bibnamefont
  {Haule}}\ and\ \bibinfo {author} {\bibfnamefont {K.}~\bibnamefont {Chen}},\
  }\bibfield  {title} {\bibinfo {title} {Single-particle excitations in the
  uniform electron gas by diagrammatic monte carlo},\ }\href
  {https://doi.org/10.1038/s41598-022-06188-6} {\bibfield  {journal} {\bibinfo
  {journal} {Scientific Reports}\ }\textbf {\bibinfo {volume} {12}},\ \bibinfo
  {pages} {2294} (\bibinfo {year} {2022})}\BibitemShut {NoStop}%
\bibitem [{\citenamefont {Gori-Giorgi}\ and\ \citenamefont
  {Ziesche}(2002)}]{PhysRevB.66.235116}%
  \BibitemOpen
  \bibfield  {author} {\bibinfo {author} {\bibfnamefont {P.}~\bibnamefont
  {Gori-Giorgi}}\ and\ \bibinfo {author} {\bibfnamefont {P.}~\bibnamefont
  {Ziesche}},\ }\bibfield  {title} {\bibinfo {title} {Momentum distribution of
  the uniform electron gas: Improved parametrization and exact limits of the
  cumulant expansion},\ }\href {https://doi.org/10.1103/PhysRevB.66.235116}
  {\bibfield  {journal} {\bibinfo  {journal} {Phys. Rev. B}\ }\textbf {\bibinfo
  {volume} {66}},\ \bibinfo {pages} {235116} (\bibinfo {year}
  {2002})}\BibitemShut {NoStop}%
\bibitem [{\citenamefont {Vogt}\ \emph {et~al.}(2004)\citenamefont {Vogt},
  \citenamefont {Zimmermann},\ and\ \citenamefont
  {Needs}}]{PhysRevB.69.045113}%
  \BibitemOpen
  \bibfield  {author} {\bibinfo {author} {\bibfnamefont {M.}~\bibnamefont
  {Vogt}}, \bibinfo {author} {\bibfnamefont {R.}~\bibnamefont {Zimmermann}},\
  and\ \bibinfo {author} {\bibfnamefont {R.~J.}\ \bibnamefont {Needs}},\
  }\bibfield  {title} {\bibinfo {title} {Spectral moments in the homogeneous
  electron gas},\ }\href {https://doi.org/10.1103/PhysRevB.69.045113}
  {\bibfield  {journal} {\bibinfo  {journal} {Phys. Rev. B}\ }\textbf {\bibinfo
  {volume} {69}},\ \bibinfo {pages} {045113} (\bibinfo {year}
  {2004})}\BibitemShut {NoStop}%
\bibitem [{\citenamefont {Georges}\ \emph {et~al.}(1996)\citenamefont
  {Georges}, \citenamefont {Kotliar}, \citenamefont {Krauth},\ and\
  \citenamefont {Rozenberg}}]{rmp_dmft}%
  \BibitemOpen
  \bibfield  {author} {\bibinfo {author} {\bibfnamefont {A.}~\bibnamefont
  {Georges}}, \bibinfo {author} {\bibfnamefont {G.}~\bibnamefont {Kotliar}},
  \bibinfo {author} {\bibfnamefont {W.}~\bibnamefont {Krauth}},\ and\ \bibinfo
  {author} {\bibfnamefont {M.~J.}\ \bibnamefont {Rozenberg}},\ }\bibfield
  {title} {\bibinfo {title} {Dynamical mean-field theory of strongly correlated
  fermion systems and the limit of infinite dimensions},\ }\href
  {https://doi.org/10.1103/RevModPhys.68.13} {\bibfield  {journal} {\bibinfo
  {journal} {Rev. Mod. Phys.}\ }\textbf {\bibinfo {volume} {68}},\ \bibinfo
  {pages} {13} (\bibinfo {year} {1996})}\BibitemShut {NoStop}%
\bibitem [{\citenamefont {Kotliar}\ \emph {et~al.}(2006)\citenamefont
  {Kotliar}, \citenamefont {Savrasov}, \citenamefont {Haule}, \citenamefont
  {Oudovenko}, \citenamefont {Parcollet},\ and\ \citenamefont
  {Marianetti}}]{RevModPhys.78.865}%
  \BibitemOpen
  \bibfield  {author} {\bibinfo {author} {\bibfnamefont {G.}~\bibnamefont
  {Kotliar}}, \bibinfo {author} {\bibfnamefont {S.~Y.}\ \bibnamefont
  {Savrasov}}, \bibinfo {author} {\bibfnamefont {K.}~\bibnamefont {Haule}},
  \bibinfo {author} {\bibfnamefont {V.~S.}\ \bibnamefont {Oudovenko}}, \bibinfo
  {author} {\bibfnamefont {O.}~\bibnamefont {Parcollet}},\ and\ \bibinfo
  {author} {\bibfnamefont {C.~A.}\ \bibnamefont {Marianetti}},\ }\bibfield
  {title} {\bibinfo {title} {Electronic structure calculations with dynamical
  mean-field theory},\ }\href {https://doi.org/10.1103/RevModPhys.78.865}
  {\bibfield  {journal} {\bibinfo  {journal} {Rev. Mod. Phys.}\ }\textbf
  {\bibinfo {volume} {78}},\ \bibinfo {pages} {865} (\bibinfo {year}
  {2006})}\BibitemShut {NoStop}%
\bibitem [{\citenamefont {Krien}\ \emph {et~al.}(2019)\citenamefont {Krien},
  \citenamefont {van Loon}, \citenamefont {Katsnelson}, \citenamefont
  {Lichtenstein},\ and\ \citenamefont {Capone}}]{PhysRevB.99.245128}%
  \BibitemOpen
  \bibfield  {author} {\bibinfo {author} {\bibfnamefont {F.}~\bibnamefont
  {Krien}}, \bibinfo {author} {\bibfnamefont {E.~G. C.~P.}\ \bibnamefont {van
  Loon}}, \bibinfo {author} {\bibfnamefont {M.~I.}\ \bibnamefont {Katsnelson}},
  \bibinfo {author} {\bibfnamefont {A.~I.}\ \bibnamefont {Lichtenstein}},\ and\
  \bibinfo {author} {\bibfnamefont {M.}~\bibnamefont {Capone}},\ }\bibfield
  {title} {\bibinfo {title} {Two-particle fermi liquid parameters at the mott
  transition: Vertex divergences, landau parameters, and incoherent response in
  dynamical mean-field theory},\ }\href
  {https://doi.org/10.1103/PhysRevB.99.245128} {\bibfield  {journal} {\bibinfo
  {journal} {Phys. Rev. B}\ }\textbf {\bibinfo {volume} {99}},\ \bibinfo
  {pages} {245128} (\bibinfo {year} {2019})}\BibitemShut {NoStop}%
\bibitem [{\citenamefont {Hohenberg}\ and\ \citenamefont
  {Kohn}(1964)}]{PhysRev.136.B864}%
  \BibitemOpen
  \bibfield  {author} {\bibinfo {author} {\bibfnamefont {P.}~\bibnamefont
  {Hohenberg}}\ and\ \bibinfo {author} {\bibfnamefont {W.}~\bibnamefont
  {Kohn}},\ }\bibfield  {title} {\bibinfo {title} {Inhomogeneous electron
  gas},\ }\href {https://doi.org/10.1103/PhysRev.136.B864} {\bibfield
  {journal} {\bibinfo  {journal} {Phys. Rev.}\ }\textbf {\bibinfo {volume}
  {136}},\ \bibinfo {pages} {B864} (\bibinfo {year} {1964})}\BibitemShut
  {NoStop}%
\bibitem [{\citenamefont {Kohn}\ and\ \citenamefont
  {Sham}(1965)}]{PhysRev.140.A1133}%
  \BibitemOpen
  \bibfield  {author} {\bibinfo {author} {\bibfnamefont {W.}~\bibnamefont
  {Kohn}}\ and\ \bibinfo {author} {\bibfnamefont {L.~J.}\ \bibnamefont
  {Sham}},\ }\bibfield  {title} {\bibinfo {title} {Self-consistent equations
  including exchange and correlation effects},\ }\href
  {https://doi.org/10.1103/PhysRev.140.A1133} {\bibfield  {journal} {\bibinfo
  {journal} {Phys. Rev.}\ }\textbf {\bibinfo {volume} {140}},\ \bibinfo {pages}
  {A1133} (\bibinfo {year} {1965})}\BibitemShut {NoStop}%
\bibitem [{\citenamefont {Mandal}\ \emph {et~al.}(2022)\citenamefont {Mandal},
  \citenamefont {Haule}, \citenamefont {Rabe},\ and\ \citenamefont
  {Vanderbilt}}]{Mandal2022}%
  \BibitemOpen
  \bibfield  {author} {\bibinfo {author} {\bibfnamefont {S.}~\bibnamefont
  {Mandal}}, \bibinfo {author} {\bibfnamefont {K.}~\bibnamefont {Haule}},
  \bibinfo {author} {\bibfnamefont {K.~M.}\ \bibnamefont {Rabe}},\ and\
  \bibinfo {author} {\bibfnamefont {D.}~\bibnamefont {Vanderbilt}},\ }\bibfield
   {title} {\bibinfo {title} {Electronic correlation in nearly free electron
  metals with beyond-dft methods},\ }\href
  {https://doi.org/10.1038/s41524-022-00867-8} {\bibfield  {journal} {\bibinfo
  {journal} {npj Computational Materials}\ }\textbf {\bibinfo {volume} {8}},\
  \bibinfo {pages} {181} (\bibinfo {year} {2022})}\BibitemShut {NoStop}%
\bibitem [{\citenamefont {Mahan}\ and\ \citenamefont
  {Sernelius}(1989)}]{PhysRevLett.62.2718}%
  \BibitemOpen
  \bibfield  {author} {\bibinfo {author} {\bibfnamefont {G.~D.}\ \bibnamefont
  {Mahan}}\ and\ \bibinfo {author} {\bibfnamefont {B.~E.}\ \bibnamefont
  {Sernelius}},\ }\bibfield  {title} {\bibinfo {title} {Electron-electron
  interactions and the bandwidth of metals},\ }\href
  {https://doi.org/10.1103/PhysRevLett.62.2718} {\bibfield  {journal} {\bibinfo
   {journal} {Phys. Rev. Lett.}\ }\textbf {\bibinfo {volume} {62}},\ \bibinfo
  {pages} {2718} (\bibinfo {year} {1989})}\BibitemShut {NoStop}%
\bibitem [{\citenamefont {Freimuth}\ \emph
  {et~al.}(2022{\natexlab{a}})\citenamefont {Freimuth}, \citenamefont
  {Bl\"ugel},\ and\ \citenamefont {Mokrousov}}]{momentis}%
  \BibitemOpen
  \bibfield  {author} {\bibinfo {author} {\bibfnamefont {F.}~\bibnamefont
  {Freimuth}}, \bibinfo {author} {\bibfnamefont {S.}~\bibnamefont {Bl\"ugel}},\
  and\ \bibinfo {author} {\bibfnamefont {Y.}~\bibnamefont {Mokrousov}},\
  }\bibfield  {title} {\bibinfo {title} {Moment functional based spectral
  density functional theory},\ }\href
  {https://doi.org/10.1103/PhysRevB.106.155114} {\bibfield  {journal} {\bibinfo
   {journal} {Phys. Rev. B}\ }\textbf {\bibinfo {volume} {106}},\ \bibinfo
  {pages} {155114} (\bibinfo {year} {2022}{\natexlab{a}})}\BibitemShut
  {NoStop}%
\bibitem [{\citenamefont {Shick}\ \emph {et~al.}(1999)\citenamefont {Shick},
  \citenamefont {Liechtenstein},\ and\ \citenamefont
  {Pickett}}]{PhysRevB.60.10763}%
  \BibitemOpen
  \bibfield  {author} {\bibinfo {author} {\bibfnamefont {A.~B.}\ \bibnamefont
  {Shick}}, \bibinfo {author} {\bibfnamefont {A.~I.}\ \bibnamefont
  {Liechtenstein}},\ and\ \bibinfo {author} {\bibfnamefont {W.~E.}\
  \bibnamefont {Pickett}},\ }\bibfield  {title} {\bibinfo {title}
  {Implementation of the lda+u method using the full-potential linearized
  augmented plane-wave basis},\ }\href
  {https://doi.org/10.1103/PhysRevB.60.10763} {\bibfield  {journal} {\bibinfo
  {journal} {Phys. Rev. B}\ }\textbf {\bibinfo {volume} {60}},\ \bibinfo
  {pages} {10763} (\bibinfo {year} {1999})}\BibitemShut {NoStop}%
\bibitem [{\citenamefont {Li}\ \emph {et~al.}(1990)\citenamefont {Li},
  \citenamefont {Freeman}, \citenamefont {Jansen},\ and\ \citenamefont
  {Fu}}]{second_variation_soi}%
  \BibitemOpen
  \bibfield  {author} {\bibinfo {author} {\bibfnamefont {C.}~\bibnamefont
  {Li}}, \bibinfo {author} {\bibfnamefont {A.~J.}\ \bibnamefont {Freeman}},
  \bibinfo {author} {\bibfnamefont {H.~J.~F.}\ \bibnamefont {Jansen}},\ and\
  \bibinfo {author} {\bibfnamefont {C.~L.}\ \bibnamefont {Fu}},\ }\bibfield
  {title} {\bibinfo {title} {Magnetic anisotropy in low-dimensional
  ferromagnetic systems: Fe monolayers on ag(001), au(001), and pd(001)
  substrates},\ }\href@noop {} {\bibfield  {journal} {\bibinfo  {journal}
  {Phys. Rev. B}\ }\textbf {\bibinfo {volume} {42}},\ \bibinfo {pages} {5433}
  (\bibinfo {year} {1990})}\BibitemShut {NoStop}%
\bibitem [{\citenamefont {Freimuth}\ \emph
  {et~al.}(2022{\natexlab{b}})\citenamefont {Freimuth}, \citenamefont
  {Blügel},\ and\ \citenamefont {Mokrousov}}]{momdis}%
  \BibitemOpen
  \bibfield  {author} {\bibinfo {author} {\bibfnamefont {F.}~\bibnamefont
  {Freimuth}}, \bibinfo {author} {\bibfnamefont {S.}~\bibnamefont {Blügel}},\
  and\ \bibinfo {author} {\bibfnamefont {Y.}~\bibnamefont {Mokrousov}},\
  }\href@noop {} {\bibinfo {title} {Moment potentials for spectral density
  functional theory: Exploiting the momentum distribution of the uniform
  electron gas}} (\bibinfo {year} {2022}{\natexlab{b}}),\ \Eprint
  {https://arxiv.org/abs/2212.12624v1} {arXiv:2212.12624v1 [cond-mat.mtrl-sci]}
  \BibitemShut {NoStop}%
\bibitem [{\citenamefont {Mahan}(2000)}]{book_mahan}%
  \BibitemOpen
  \bibfield  {author} {\bibinfo {author} {\bibfnamefont {G.~D.}\ \bibnamefont
  {Mahan}},\ }\href@noop {} {\emph {\bibinfo {title} {Many-Particle
  Physics}}},\ Physics of Solids and Liquids\ (\bibinfo  {publisher} {Kluwer
  Academic/Plenum Publishers},\ \bibinfo {year} {2000})\BibitemShut {NoStop}%
\bibitem [{\citenamefont {Martin}(2020)}]{book_martin_2020}%
  \BibitemOpen
  \bibfield  {author} {\bibinfo {author} {\bibfnamefont {R.~M.}\ \bibnamefont
  {Martin}},\ }\href@noop {} {\emph {\bibinfo {title} {Electronic Structure:
  Basic Theory and Practical Methods}}},\ \bibinfo {edition} {2nd}\ ed.\
  (\bibinfo  {publisher} {Cambridge University Press},\ \bibinfo {year}
  {2020})\BibitemShut {NoStop}%
\bibitem [{\citenamefont {van Schilfgaarde}\ \emph {et~al.}(2006)\citenamefont
  {van Schilfgaarde}, \citenamefont {Kotani},\ and\ \citenamefont
  {Faleev}}]{PhysRevLett.96.226402}%
  \BibitemOpen
  \bibfield  {author} {\bibinfo {author} {\bibfnamefont {M.}~\bibnamefont {van
  Schilfgaarde}}, \bibinfo {author} {\bibfnamefont {T.}~\bibnamefont
  {Kotani}},\ and\ \bibinfo {author} {\bibfnamefont {S.}~\bibnamefont
  {Faleev}},\ }\bibfield  {title} {\bibinfo {title} {Quasiparticle
  self-consistent $gw$ theory},\ }\href
  {https://doi.org/10.1103/PhysRevLett.96.226402} {\bibfield  {journal}
  {\bibinfo  {journal} {Phys. Rev. Lett.}\ }\textbf {\bibinfo {volume} {96}},\
  \bibinfo {pages} {226402} (\bibinfo {year} {2006})}\BibitemShut {NoStop}%
\bibitem [{\citenamefont {Sekiyama}\ \emph {et~al.}(2004)\citenamefont
  {Sekiyama}, \citenamefont {Fujiwara}, \citenamefont {Imada}, \citenamefont
  {Suga}, \citenamefont {Eisaki}, \citenamefont {Uchida}, \citenamefont
  {Takegahara}, \citenamefont {Harima}, \citenamefont {Saitoh}, \citenamefont
  {Nekrasov}, \citenamefont {Keller}, \citenamefont {Kondakov}, \citenamefont
  {Kozhevnikov}, \citenamefont {Pruschke}, \citenamefont {Held}, \citenamefont
  {Vollhardt},\ and\ \citenamefont {Anisimov}}]{PhysRevLett.93.156402}%
  \BibitemOpen
  \bibfield  {author} {\bibinfo {author} {\bibfnamefont {A.}~\bibnamefont
  {Sekiyama}}, \bibinfo {author} {\bibfnamefont {H.}~\bibnamefont {Fujiwara}},
  \bibinfo {author} {\bibfnamefont {S.}~\bibnamefont {Imada}}, \bibinfo
  {author} {\bibfnamefont {S.}~\bibnamefont {Suga}}, \bibinfo {author}
  {\bibfnamefont {H.}~\bibnamefont {Eisaki}}, \bibinfo {author} {\bibfnamefont
  {S.~I.}\ \bibnamefont {Uchida}}, \bibinfo {author} {\bibfnamefont
  {K.}~\bibnamefont {Takegahara}}, \bibinfo {author} {\bibfnamefont
  {H.}~\bibnamefont {Harima}}, \bibinfo {author} {\bibfnamefont
  {Y.}~\bibnamefont {Saitoh}}, \bibinfo {author} {\bibfnamefont {I.~A.}\
  \bibnamefont {Nekrasov}}, \bibinfo {author} {\bibfnamefont {G.}~\bibnamefont
  {Keller}}, \bibinfo {author} {\bibfnamefont {D.~E.}\ \bibnamefont
  {Kondakov}}, \bibinfo {author} {\bibfnamefont {A.~V.}\ \bibnamefont
  {Kozhevnikov}}, \bibinfo {author} {\bibfnamefont {T.}~\bibnamefont
  {Pruschke}}, \bibinfo {author} {\bibfnamefont {K.}~\bibnamefont {Held}},
  \bibinfo {author} {\bibfnamefont {D.}~\bibnamefont {Vollhardt}},\ and\
  \bibinfo {author} {\bibfnamefont {V.~I.}\ \bibnamefont {Anisimov}},\
  }\bibfield  {title} {\bibinfo {title} {Mutual experimental and theoretical
  validation of bulk photoemission spectra of
  ${\mathrm{s}\mathrm{r}}_{1\ensuremath{-}x}{\mathrm{c}\mathrm{a}}_{x}{\mathrm{v}\mathrm{o}}_{3}$},\
  }\href {https://doi.org/10.1103/PhysRevLett.93.156402} {\bibfield  {journal}
  {\bibinfo  {journal} {Phys. Rev. Lett.}\ }\textbf {\bibinfo {volume} {93}},\
  \bibinfo {pages} {156402} (\bibinfo {year} {2004})}\BibitemShut {NoStop}%
\bibitem [{\citenamefont {Nekrasov}\ \emph {et~al.}(2005)\citenamefont
  {Nekrasov}, \citenamefont {Keller}, \citenamefont {Kondakov}, \citenamefont
  {Kozhevnikov}, \citenamefont {Pruschke}, \citenamefont {Held}, \citenamefont
  {Vollhardt},\ and\ \citenamefont {Anisimov}}]{PhysRevB.72.155106}%
  \BibitemOpen
  \bibfield  {author} {\bibinfo {author} {\bibfnamefont {I.~A.}\ \bibnamefont
  {Nekrasov}}, \bibinfo {author} {\bibfnamefont {G.}~\bibnamefont {Keller}},
  \bibinfo {author} {\bibfnamefont {D.~E.}\ \bibnamefont {Kondakov}}, \bibinfo
  {author} {\bibfnamefont {A.~V.}\ \bibnamefont {Kozhevnikov}}, \bibinfo
  {author} {\bibfnamefont {T.}~\bibnamefont {Pruschke}}, \bibinfo {author}
  {\bibfnamefont {K.}~\bibnamefont {Held}}, \bibinfo {author} {\bibfnamefont
  {D.}~\bibnamefont {Vollhardt}},\ and\ \bibinfo {author} {\bibfnamefont
  {V.~I.}\ \bibnamefont {Anisimov}},\ }\bibfield  {title} {\bibinfo {title}
  {Comparative study of correlation effects in
  $\mathrm{Ca}\mathrm{V}{\mathrm{o}}_{3}$ and
  $\mathrm{Sr}\mathrm{V}{\mathrm{o}}_{3}$},\ }\href
  {https://doi.org/10.1103/PhysRevB.72.155106} {\bibfield  {journal} {\bibinfo
  {journal} {Phys. Rev. B}\ }\textbf {\bibinfo {volume} {72}},\ \bibinfo
  {pages} {155106} (\bibinfo {year} {2005})}\BibitemShut {NoStop}%
\bibitem [{\citenamefont {Nolting}\ and\ \citenamefont
  {Brewer}(2009)}]{book_Nolting}%
  \BibitemOpen
  \bibfield  {author} {\bibinfo {author} {\bibfnamefont {W.}~\bibnamefont
  {Nolting}}\ and\ \bibinfo {author} {\bibfnamefont {W.}~\bibnamefont
  {Brewer}},\ }\href@noop {} {\emph {\bibinfo {title} {Fundamentals of
  Many-body Physics: Principles and Methods}}}\ (\bibinfo  {publisher}
  {Springer Berlin Heidelberg},\ \bibinfo {year} {2009})\BibitemShut {NoStop}%
\bibitem [{\citenamefont {Nolting}\ \emph {et~al.}(1989)\citenamefont
  {Nolting}, \citenamefont {Borgiel}, \citenamefont {Dose},\ and\ \citenamefont
  {Fauster}}]{nickel_PhysRevB.40.5015}%
  \BibitemOpen
  \bibfield  {author} {\bibinfo {author} {\bibfnamefont {W.}~\bibnamefont
  {Nolting}}, \bibinfo {author} {\bibfnamefont {W.}~\bibnamefont {Borgiel}},
  \bibinfo {author} {\bibfnamefont {V.}~\bibnamefont {Dose}},\ and\ \bibinfo
  {author} {\bibfnamefont {T.}~\bibnamefont {Fauster}},\ }\bibfield  {title}
  {\bibinfo {title} {Finite-temperature ferromagnetism of nickel},\ }\href
  {https://doi.org/10.1103/PhysRevB.40.5015} {\bibfield  {journal} {\bibinfo
  {journal} {Phys. Rev. B}\ }\textbf {\bibinfo {volume} {40}},\ \bibinfo
  {pages} {5015} (\bibinfo {year} {1989})}\BibitemShut {NoStop}%
\bibitem [{\citenamefont {Borgiel}\ and\ \citenamefont
  {Nolting}(1990)}]{nickel_Borgiel1990}%
  \BibitemOpen
  \bibfield  {author} {\bibinfo {author} {\bibfnamefont {W.}~\bibnamefont
  {Borgiel}}\ and\ \bibinfo {author} {\bibfnamefont {W.}~\bibnamefont
  {Nolting}},\ }\bibfield  {title} {\bibinfo {title} {Many body contributions
  to the electronic structure of nickel},\ }\href
  {https://doi.org/10.1007/BF01307842} {\bibfield  {journal} {\bibinfo
  {journal} {Zeitschrift f{\"u}r Physik B Condensed Matter}\ }\textbf {\bibinfo
  {volume} {78}},\ \bibinfo {pages} {241} (\bibinfo {year} {1990})}\BibitemShut
  {NoStop}%
\bibitem [{\citenamefont {Nolting}\ and\ \citenamefont
  {Oles}(1980)}]{Nolting_1980}%
  \BibitemOpen
  \bibfield  {author} {\bibinfo {author} {\bibfnamefont {W.}~\bibnamefont
  {Nolting}}\ and\ \bibinfo {author} {\bibfnamefont {A.~M.}\ \bibnamefont
  {Oles}},\ }\bibfield  {title} {\bibinfo {title} {Spectral density approach
  for the quasiparticle concept in the s-f model (ferromagnetic
  semiconductors)},\ }\href {https://doi.org/10.1088/0022-3719/13/12/012}
  {\bibfield  {journal} {\bibinfo  {journal} {Journal of Physics C: Solid State
  Physics}\ }\textbf {\bibinfo {volume} {13}},\ \bibinfo {pages} {2295}
  (\bibinfo {year} {1980})}\BibitemShut {NoStop}%
\bibitem [{\citenamefont {Ortiz}\ and\ \citenamefont
  {Ballone}(1994)}]{PhysRevB.50.1391}%
  \BibitemOpen
  \bibfield  {author} {\bibinfo {author} {\bibfnamefont {G.}~\bibnamefont
  {Ortiz}}\ and\ \bibinfo {author} {\bibfnamefont {P.}~\bibnamefont
  {Ballone}},\ }\bibfield  {title} {\bibinfo {title} {Correlation energy,
  structure factor, radial distribution function, and momentum distribution of
  the spin-polarized uniform electron gas},\ }\href
  {https://doi.org/10.1103/PhysRevB.50.1391} {\bibfield  {journal} {\bibinfo
  {journal} {Phys. Rev. B}\ }\textbf {\bibinfo {volume} {50}},\ \bibinfo
  {pages} {1391} (\bibinfo {year} {1994})}\BibitemShut {NoStop}%
\bibitem [{\citenamefont {Ortiz}\ and\ \citenamefont
  {Ballone}(1997)}]{PhysRevB.56.9970}%
  \BibitemOpen
  \bibfield  {author} {\bibinfo {author} {\bibfnamefont {G.}~\bibnamefont
  {Ortiz}}\ and\ \bibinfo {author} {\bibfnamefont {P.}~\bibnamefont
  {Ballone}},\ }\bibfield  {title} {\bibinfo {title} {Erratum: Correlation
  energy, structure factor, radial distribution function, and momentum
  distribution of the spin-polarized uniform electron gas [phys. rev. b 50,
  1391 (1994)]},\ }\href {https://doi.org/10.1103/PhysRevB.56.9970} {\bibfield
  {journal} {\bibinfo  {journal} {Phys. Rev. B}\ }\textbf {\bibinfo {volume}
  {56}},\ \bibinfo {pages} {9970(E)} (\bibinfo {year} {1997})}\BibitemShut
  {NoStop}%
\bibitem [{\citenamefont {Giuliani}\ and\ \citenamefont
  {Vignale}(2005)}]{giuliani_vignale_2005}%
  \BibitemOpen
  \bibfield  {author} {\bibinfo {author} {\bibfnamefont {G.}~\bibnamefont
  {Giuliani}}\ and\ \bibinfo {author} {\bibfnamefont {G.}~\bibnamefont
  {Vignale}},\ }\href {https://doi.org/10.1017/CBO9780511619915} {\emph
  {\bibinfo {title} {Quantum Theory of the Electron Liquid}}}\ (\bibinfo
  {publisher} {Cambridge University Press},\ \bibinfo {year}
  {2005})\BibitemShut {NoStop}%
\bibitem [{\citenamefont {Chen}\ and\ \citenamefont
  {Haule}(2019)}]{vdmc_Chen2019}%
  \BibitemOpen
  \bibfield  {author} {\bibinfo {author} {\bibfnamefont {K.}~\bibnamefont
  {Chen}}\ and\ \bibinfo {author} {\bibfnamefont {K.}~\bibnamefont {Haule}},\
  }\bibfield  {title} {\bibinfo {title} {A combined variational and
  diagrammatic quantum monte carlo approach to the many-electron problem},\
  }\href {https://doi.org/10.1038/s41467-019-11708-6} {\bibfield  {journal}
  {\bibinfo  {journal} {Nature Communications}\ }\textbf {\bibinfo {volume}
  {10}},\ \bibinfo {pages} {3725} (\bibinfo {year} {2019})}\BibitemShut
  {NoStop}%
\bibitem [{\citenamefont {Holzmann}\ \emph {et~al.}(2011)\citenamefont
  {Holzmann}, \citenamefont {Bernu}, \citenamefont {Pierleoni}, \citenamefont
  {McMinis}, \citenamefont {Ceperley}, \citenamefont {Olevano},\ and\
  \citenamefont {Delle~Site}}]{PhysRevLett.107.110402}%
  \BibitemOpen
  \bibfield  {author} {\bibinfo {author} {\bibfnamefont {M.}~\bibnamefont
  {Holzmann}}, \bibinfo {author} {\bibfnamefont {B.}~\bibnamefont {Bernu}},
  \bibinfo {author} {\bibfnamefont {C.}~\bibnamefont {Pierleoni}}, \bibinfo
  {author} {\bibfnamefont {J.}~\bibnamefont {McMinis}}, \bibinfo {author}
  {\bibfnamefont {D.~M.}\ \bibnamefont {Ceperley}}, \bibinfo {author}
  {\bibfnamefont {V.}~\bibnamefont {Olevano}},\ and\ \bibinfo {author}
  {\bibfnamefont {L.}~\bibnamefont {Delle~Site}},\ }\bibfield  {title}
  {\bibinfo {title} {Momentum distribution of the homogeneous electron gas},\
  }\href {https://doi.org/10.1103/PhysRevLett.107.110402} {\bibfield  {journal}
  {\bibinfo  {journal} {Phys. Rev. Lett.}\ }\textbf {\bibinfo {volume} {107}},\
  \bibinfo {pages} {110402} (\bibinfo {year} {2011})}\BibitemShut {NoStop}%
\bibitem [{\citenamefont {Perdew}\ \emph {et~al.}(1996)\citenamefont {Perdew},
  \citenamefont {Burke},\ and\ \citenamefont
  {Ernzerhof}}]{PhysRevLett.77.3865}%
  \BibitemOpen
  \bibfield  {author} {\bibinfo {author} {\bibfnamefont {J.~P.}\ \bibnamefont
  {Perdew}}, \bibinfo {author} {\bibfnamefont {K.}~\bibnamefont {Burke}},\ and\
  \bibinfo {author} {\bibfnamefont {M.}~\bibnamefont {Ernzerhof}},\ }\bibfield
  {title} {\bibinfo {title} {Generalized gradient approximation made simple},\
  }\href {https://doi.org/10.1103/PhysRevLett.77.3865} {\bibfield  {journal}
  {\bibinfo  {journal} {Phys. Rev. Lett.}\ }\textbf {\bibinfo {volume} {77}},\
  \bibinfo {pages} {3865} (\bibinfo {year} {1996})}\BibitemShut {NoStop}%
\bibitem [{\citenamefont {Cassabois}\ \emph {et~al.}(2016)\citenamefont
  {Cassabois}, \citenamefont {Valvin},\ and\ \citenamefont
  {Gil}}]{Cassabois2016}%
  \BibitemOpen
  \bibfield  {author} {\bibinfo {author} {\bibfnamefont {G.}~\bibnamefont
  {Cassabois}}, \bibinfo {author} {\bibfnamefont {P.}~\bibnamefont {Valvin}},\
  and\ \bibinfo {author} {\bibfnamefont {B.}~\bibnamefont {Gil}},\ }\bibfield
  {title} {\bibinfo {title} {Hexagonal boron nitride is an indirect bandgap
  semiconductor},\ }\href {https://doi.org/10.1038/nphoton.2015.277} {\bibfield
   {journal} {\bibinfo  {journal} {Nature Photonics}\ }\textbf {\bibinfo
  {volume} {10}},\ \bibinfo {pages} {262} (\bibinfo {year} {2016})}\BibitemShut
  {NoStop}%
\bibitem [{\citenamefont {Nolting}\ \emph {et~al.}(1995)\citenamefont
  {Nolting}, \citenamefont {Vega},\ and\ \citenamefont
  {Fauster}}]{bcc_iron_Nolting1995}%
  \BibitemOpen
  \bibfield  {author} {\bibinfo {author} {\bibfnamefont {W.}~\bibnamefont
  {Nolting}}, \bibinfo {author} {\bibfnamefont {A.}~\bibnamefont {Vega}},\ and\
  \bibinfo {author} {\bibfnamefont {T.}~\bibnamefont {Fauster}},\ }\bibfield
  {title} {\bibinfo {title} {Electronic quasiparticle structure of
  ferromagnetic bcc iron},\ }\href {https://doi.org/10.1007/BF01313058}
  {\bibfield  {journal} {\bibinfo  {journal} {Zeitschrift f{\"u}r Physik B
  Condensed Matter}\ }\textbf {\bibinfo {volume} {96}},\ \bibinfo {pages} {357}
  (\bibinfo {year} {1995})}\BibitemShut {NoStop}%
\bibitem [{\citenamefont {Raaen}\ and\ \citenamefont
  {Murgai}(1987)}]{PhysRevB.36.887}%
  \BibitemOpen
  \bibfield  {author} {\bibinfo {author} {\bibfnamefont {S.}~\bibnamefont
  {Raaen}}\ and\ \bibinfo {author} {\bibfnamefont {V.}~\bibnamefont {Murgai}},\
  }\bibfield  {title} {\bibinfo {title} {Absence of two-electron resonances in
  valence-band photoemission from cr, mn, fe, and co},\ }\href
  {https://doi.org/10.1103/PhysRevB.36.887} {\bibfield  {journal} {\bibinfo
  {journal} {Phys. Rev. B}\ }\textbf {\bibinfo {volume} {36}},\ \bibinfo
  {pages} {887} (\bibinfo {year} {1987})}\BibitemShut {NoStop}%
\bibitem [{\citenamefont {Kakizaki}\ \emph {et~al.}(1997)\citenamefont
  {Kakizaki}, \citenamefont {Ono}, \citenamefont {Tanaka}, \citenamefont
  {Shimada},\ and\ \citenamefont {Sendohda}}]{PhysRevB.55.6678}%
  \BibitemOpen
  \bibfield  {author} {\bibinfo {author} {\bibfnamefont {A.}~\bibnamefont
  {Kakizaki}}, \bibinfo {author} {\bibfnamefont {K.}~\bibnamefont {Ono}},
  \bibinfo {author} {\bibfnamefont {K.}~\bibnamefont {Tanaka}}, \bibinfo
  {author} {\bibfnamefont {K.}~\bibnamefont {Shimada}},\ and\ \bibinfo {author}
  {\bibfnamefont {T.}~\bibnamefont {Sendohda}},\ }\bibfield  {title} {\bibinfo
  {title} {Spin-resolved photoemission of valence-band satellites of ni},\
  }\href {https://doi.org/10.1103/PhysRevB.55.6678} {\bibfield  {journal}
  {\bibinfo  {journal} {Phys. Rev. B}\ }\textbf {\bibinfo {volume} {55}},\
  \bibinfo {pages} {6678} (\bibinfo {year} {1997})}\BibitemShut {NoStop}%
\bibitem [{\citenamefont {H\"ufner}\ \emph {et~al.}(2000)\citenamefont
  {H\"ufner}, \citenamefont {Yang}, \citenamefont {Mun}, \citenamefont
  {Fadley}, \citenamefont {Sch\"afer}, \citenamefont {Rotenberg},\ and\
  \citenamefont {Kevan}}]{PhysRevB.61.12582}%
  \BibitemOpen
  \bibfield  {author} {\bibinfo {author} {\bibfnamefont {S.}~\bibnamefont
  {H\"ufner}}, \bibinfo {author} {\bibfnamefont {S.-H.}\ \bibnamefont {Yang}},
  \bibinfo {author} {\bibfnamefont {B.~S.}\ \bibnamefont {Mun}}, \bibinfo
  {author} {\bibfnamefont {C.~S.}\ \bibnamefont {Fadley}}, \bibinfo {author}
  {\bibfnamefont {J.}~\bibnamefont {Sch\"afer}}, \bibinfo {author}
  {\bibfnamefont {E.}~\bibnamefont {Rotenberg}},\ and\ \bibinfo {author}
  {\bibfnamefont {S.~D.}\ \bibnamefont {Kevan}},\ }\bibfield  {title} {\bibinfo
  {title} {Observation of the two-hole satellite in cr and fe metal by resonant
  photoemission at the $2p$ absorption energy},\ }\href
  {https://doi.org/10.1103/PhysRevB.61.12582} {\bibfield  {journal} {\bibinfo
  {journal} {Phys. Rev. B}\ }\textbf {\bibinfo {volume} {61}},\ \bibinfo
  {pages} {12582} (\bibinfo {year} {2000})}\BibitemShut {NoStop}%
\bibitem [{\citenamefont {Nakajima}\ \emph {et~al.}(2004)\citenamefont
  {Nakajima}, \citenamefont {Hatta}, \citenamefont {Odagiri}, \citenamefont
  {Kato},\ and\ \citenamefont {Sakisaka}}]{PhysRevB.70.233103}%
  \BibitemOpen
  \bibfield  {author} {\bibinfo {author} {\bibfnamefont {N.}~\bibnamefont
  {Nakajima}}, \bibinfo {author} {\bibfnamefont {S.}~\bibnamefont {Hatta}},
  \bibinfo {author} {\bibfnamefont {J.}~\bibnamefont {Odagiri}}, \bibinfo
  {author} {\bibfnamefont {H.}~\bibnamefont {Kato}},\ and\ \bibinfo {author}
  {\bibfnamefont {Y.}~\bibnamefont {Sakisaka}},\ }\bibfield  {title} {\bibinfo
  {title} {Valence-band satellites in ni: A photoelectron spectroscopic
  study},\ }\href {https://doi.org/10.1103/PhysRevB.70.233103} {\bibfield
  {journal} {\bibinfo  {journal} {Phys. Rev. B}\ }\textbf {\bibinfo {volume}
  {70}},\ \bibinfo {pages} {233103} (\bibinfo {year} {2004})}\BibitemShut
  {NoStop}%
\bibitem [{\citenamefont {S\'anchez-Barriga}\ \emph {et~al.}(2012)\citenamefont
  {S\'anchez-Barriga}, \citenamefont {Braun}, \citenamefont {Min\'ar},
  \citenamefont {Di~Marco}, \citenamefont {Varykhalov}, \citenamefont {Rader},
  \citenamefont {Boni}, \citenamefont {Bellini}, \citenamefont {Manghi},
  \citenamefont {Ebert}, \citenamefont {Katsnelson}, \citenamefont
  {Lichtenstein}, \citenamefont {Eriksson}, \citenamefont {Eberhardt},
  \citenamefont {D\"urr},\ and\ \citenamefont {Fink}}]{PhysRevB.85.205109}%
  \BibitemOpen
  \bibfield  {author} {\bibinfo {author} {\bibfnamefont {J.}~\bibnamefont
  {S\'anchez-Barriga}}, \bibinfo {author} {\bibfnamefont {J.}~\bibnamefont
  {Braun}}, \bibinfo {author} {\bibfnamefont {J.}~\bibnamefont {Min\'ar}},
  \bibinfo {author} {\bibfnamefont {I.}~\bibnamefont {Di~Marco}}, \bibinfo
  {author} {\bibfnamefont {A.}~\bibnamefont {Varykhalov}}, \bibinfo {author}
  {\bibfnamefont {O.}~\bibnamefont {Rader}}, \bibinfo {author} {\bibfnamefont
  {V.}~\bibnamefont {Boni}}, \bibinfo {author} {\bibfnamefont {V.}~\bibnamefont
  {Bellini}}, \bibinfo {author} {\bibfnamefont {F.}~\bibnamefont {Manghi}},
  \bibinfo {author} {\bibfnamefont {H.}~\bibnamefont {Ebert}}, \bibinfo
  {author} {\bibfnamefont {M.~I.}\ \bibnamefont {Katsnelson}}, \bibinfo
  {author} {\bibfnamefont {A.~I.}\ \bibnamefont {Lichtenstein}}, \bibinfo
  {author} {\bibfnamefont {O.}~\bibnamefont {Eriksson}}, \bibinfo {author}
  {\bibfnamefont {W.}~\bibnamefont {Eberhardt}}, \bibinfo {author}
  {\bibfnamefont {H.~A.}\ \bibnamefont {D\"urr}},\ and\ \bibinfo {author}
  {\bibfnamefont {J.}~\bibnamefont {Fink}},\ }\bibfield  {title} {\bibinfo
  {title} {Effects of spin-dependent quasiparticle renormalization in fe, co,
  and ni photoemission spectra:an experimental and theoretical study},\ }\href
  {https://doi.org/10.1103/PhysRevB.85.205109} {\bibfield  {journal} {\bibinfo
  {journal} {Phys. Rev. B}\ }\textbf {\bibinfo {volume} {85}},\ \bibinfo
  {pages} {205109} (\bibinfo {year} {2012})}\BibitemShut {NoStop}%
\bibitem [{\citenamefont {Gori-Giorgi}\ \emph {et~al.}(2000)\citenamefont
  {Gori-Giorgi}, \citenamefont {Sacchetti},\ and\ \citenamefont
  {Bachelet}}]{PhysRevB.61.7353}%
  \BibitemOpen
  \bibfield  {author} {\bibinfo {author} {\bibfnamefont {P.}~\bibnamefont
  {Gori-Giorgi}}, \bibinfo {author} {\bibfnamefont {F.}~\bibnamefont
  {Sacchetti}},\ and\ \bibinfo {author} {\bibfnamefont {G.~B.}\ \bibnamefont
  {Bachelet}},\ }\bibfield  {title} {\bibinfo {title} {Analytic static
  structure factors and pair-correlation functions for the unpolarized
  homogeneous electron gas},\ }\href {https://doi.org/10.1103/PhysRevB.61.7353}
  {\bibfield  {journal} {\bibinfo  {journal} {Phys. Rev. B}\ }\textbf {\bibinfo
  {volume} {61}},\ \bibinfo {pages} {7353} (\bibinfo {year}
  {2000})}\BibitemShut {NoStop}%
\end{thebibliography}%

\end{document}